\documentclass[12pt,letterpaper,hyper]{JHEP3}
\usepackage[active]{srcltx} %this package is to be used for inverse search (to go from Kdvi to kile by a single click)
\usepackage[latin1]{inputenc}
\usepackage{epic,eepic}
\usepackage{amsmath,epsfig}
\usepackage{amssymb,amsfonts}
\usepackage{dsfont}
\usepackage{graphicx}
\usepackage{multirow,cite}

\title{Borcherds-Kac-Moody Symmetry of $\CN=4$ Dyons}

\preprint{TIFR/TH/08-36}
\author{
Miranda C. N. Cheng$^{~1}$ and Atish Dabholkar$^{~2, ~3}$\\

\it $^1$Jefferson Physical Laboratory, \\
\it Harvard University, Cambridge, MA 02138, USA\\

\it $^2$Department of Theoretical Physics\\
\it Tata Institute of Fundamental Research\\
\it Homi Bhabha Rd, Mumbai 400 005, India\\

\it $^3${Laboratoire de Physique Th\'eorique et Hautes Energies (LPTHE)\\
\it{Universit\'e Pierre et Marie Curie-Paris 6; CNRS UMR 7589}\\
\it{Tour 24-25, 5$^{\grave{e}me}$ \'etage, Boite 126, 4 Place Jussieu} \\
\it {75252 Paris Cedex 05, France}}\\

}

\abstract{We consider compactifications of heterotic string theory to four dimensions on CHL orbifolds of the type $T^6 /\mathbb{Z}_N$ with $\CN=4$ supersymmetry. The exact partition functions of the quarter-BPS dyons in these models are given in terms of genus-two Siegel modular forms. Only the $N=1,2,3$ models satisfy a certain finiteness condition, and in these cases one can identify a Borcherds-Kac-Moody superalgebra underlying the symmetry structure of the dyon spectrum.
We identify the real roots, and find that the corresponding Cartan matrices exhaust a known classification.
We show that the Siegel modular form satisfies the Weyl denominator identity of the algebra, which enables the determination of all root multiplicities.
Furthermore, the Weyl group determines the structure of wall-crossings and the attractor flows of the theory.
For $N> 4$, no such interpretation appears to be possible.
}

\keywords{black holes, superstrings, dyons}

\newenvironment{myenumerate}{
\begin{enumerate}
   \setlength{\itemsep}{1pt}
   \setlength{\parskip}{0pt}
   \setlength{\parsep}{0pt}}{\end{enumerate}}
\newenvironment{myitemize}{
\begin{itemize}
   \setlength{\itemsep}{1pt}
   \setlength{\parskip}{0pt}
   \setlength{\parsep}{0pt}}{\end{itemize}}
\renewcommand{\th}{\theta}
\renewcommand{\Im}{\mbox{Im}}
\renewcommand{\Re}{\mbox{Re}}

\newcommand{\pa}{\partial}

\newcommand{\Tr}{\mbox{Tr}}

\def\d{\partial}
\def\r{\rho}
\def\f{\phi}
\def\s{\sigma}
\def\g{\gamma}
\def\t{\tau}
\def\a{\alpha}
\def\b{\beta}
\def\m{\mu}
\def\n{\nu}
\def\e{\epsilon}

\def\l{{\lambda}}

\def\O{{\Omega}}
\def\D{\Delta}
\def\G{\Gamma}
\def\L{I}

\def\CN{{\cal N}}

\def\CN{{\cal N}}

\def\SS{\scriptscriptstyle}

\def\p{\partial}

\def\CN{{\cal N}}

\def\Tr{{\rm Tr}}

\font\manual=manfnt
\def\dbend{\lower3.5pt\hbox{\manual\char127}}

\def\c{\cdot}

\def\p{\partial}

\def\bar{\overline}

\def\CN{{\cal N}}

\def\rt2{\sqrt{2}}
\def\irt2{{1\over\sqrt{2}}}

\def\t{\tilde}

\def\s{\sigma}
\def\b{\beta}
\def\a{\alpha}

\def\g{\gamma}

\font\cmss=cmss10
\font\cmsss=cmss10 at 7pt
\def\IL{\relax{\rm I\kern-.18em L}}
\def\IH{\relax{\rm I\kern-.18em H}}
\def\rlx{\relax\leavevmode}
\def\ZZ{\rlx\leavevmode\ifmmode\mathchoice{\hbox{\cmss Z\kern-.4em Z}}
 {\hbox{\cmss Z\kern-.4em Z}}{\lower.9pt\hbox{\cmsss Z\kern-.36em Z}}
 {\lower1.2pt\hbox{\cmsss Z\kern-.36em Z}}\else{\cmss Z\kern-.4em
 Z}\fi}
\def\SS{\scriptscriptstyle}

\def\rt2{\sqrt{2}}
\def\irt2{{1\over\sqrt{2}}}

\def\t{\tilde}

\def\s{\sigma}

\newcommand{\Z}{{\mathbb Z}}
\newcommand{\R}{{\mathbb R}}
\newcommand{\N}{{\mathbb N}}
\newcommand{\C}{{\mathbb C}}

\newcommand{\Q}{{\mathbb Q}}

\def\Tr{{\rm Tr}}

\newcommand{\bea}{\begin{eqnarray}}
\newcommand{\eea}{\end{eqnarray}}
\newcommand{\be}{\begin{equation}}
\newcommand{\ee}{\end{equation}}
\newcommand{\ben}{\begin{eqnarray*}}
\newcommand{\een}{\end{eqnarray*}}
\newcommand{\bem}{\begin{pmatrix}}
\newcommand{\eem}{\end{pmatrix}}
\newcommand{\bl}{\begin{align}}
\newcommand{\el}{\end{align}}

\newcommand\Top{\rule{0pt}{2.6ex}}
\newcommand\Bottom{\rule[-1.2ex]{0pt}{0pt}}

\def\a{\alpha}
\def\b{\beta}
\def\c{\gamma}
\def\d{\delta}

\def\e{\epsilon}
\def\f{\phi}               %       \varphi

\def\g{\gamma}

\def\im{\mathrm{Im}}
\def\inf{\infty}

\def\l{\lambda}
\def\m{\mu}
\def\n{\nu}

\def\p{\pi}
\def\pa{\partial}
\def\re{\mathrm{Re}}                %     \vartheta
\def\r{\rho}                                     %     \varrho
\def\s{\sigma}                                   %     \varsigma
\def\t{\tau}
\def\th{\theta}
\def\til{\tilde}

\def\z{\zeta}

\def\D{\Delta}
\def\F{\Phi}
\def\G{\Gamma}

\def\L{\Lambda}
\def\O{\Omega}
\def\P{\Pi}

\begin{document}

\section{Introduction}

The spectrum of quarter-BPS dyons in various string theory compactifications with   $\CN =4$ supersymmetry in four dimensions has revealed a surprisingly rich structure \cite{Dijkgraaf:1996it,Gaiotto:2005gf,Shih:2005uc,Gaiotto:2005hc,Shih:2005he,Jatkar:2005bh,
David:2006ji,Dabholkar:2006xa,David:2006yn,David:2006ru,David:2006ud, Dabholkar:2006bj,Sen:2007vb,Dabholkar:2007vk,Cheng:2007ch}. The first surprise is of course that the \textit{exact} degeneracy of these non-perturbative states can be computed at all, especially given  the fact that these dyons are in general complicated bound states of not only D-branes but also NS5-branes, KK-monopoles, strings and momenta.  The second surprise is that the spectrum is now known for all duality orbits of  charges \cite{Dabholkar:2007vk,Banerjee:2008pv,Banerjee:2008pu,Banerjee:2008ri,Dabholkar:2008zy} at all points in the moduli space, and not just in weakly coupled corners of the moduli space. This is all the more surprising because the degeneracy of the quarter-BPS states jumps across infinitely many ``walls of marginal stability" in the moduli space. Therefore, unlike in the case of half-BPS states, one cannot naively analytically continue the weak coupling results to other regions in the moduli space\cite{Sen:2007vb,Dabholkar:2007vk,Cheng:2007ch}.

Such an exact, non-perturbative knowledge of the full quarter-BPS  spectrum at all points in the moduli space is sure to illuminate  the structure of the theory in unexpected ways. One may hope to learn something about the theory in the middle  of the moduli space where there is no weakly coupled description. Already, it has allowed us to address a number of physical questions in far greater detail than is possible for the $\CN=2$ dyons. For example, from the \textit{microscopic} analysis of the exact partition function, one can analyze very precisely the structure of the walls of marginal stability \cite{Sen:2007vb,Dabholkar:2007vk,Cheng:2007ch,Dabholkar:2008zy} and the sub-leading corrections to the  entropy of the corresponding dyonic black holes \cite{LopesCardoso:2004xf,David:2006yn}. These are found to be in beautiful agreement with the independent \textit{macroscopic} supergravity analysis \cite{LopesCardoso:1998wt,LopesCardoso:1999ur,LopesCardoso:1999xn,LopesCardoso:2004xf,
David:2006yn,Sen:2007vb,Dabholkar:2007vk,Cheng:2007ch,Banerjee:2008pv,Banerjee:2008pu,Banerjee:2008ri}.

One important motivation for studying the spectrum of dyons in such detail is to explore whether there are deeper underlying structures.  Indeed, duality symmetries in field theory were first noticed in the  dyon spectrum. In string theory, the analysis of half-BPS states was already powerful enough to suggest the intricate web of dualities. It is quite likely that the symmetries of string theory are a further generalization of known symmetries such as gauge invariance and general coordinate invariance, since duality relates different string and brane charges to each other. Exact information about the spectrum of non-perturbative dyons may suggest more concrete structures which might enable us to develop the necessary tools to come to grips with such a symmetry.

For the toroidally compactified heterotic string theory, the partition function of BPS dyons \cite{Dijkgraaf:1996it} is given in terms of the so-called `Igusa cusp form' which is the unique  weight ten Siegel modular form of the group $Sp(2, \mathbb{Z})$. It has been long known that the square root of the Igusa form equals  the denominator of a generalized (or Borcherds-) Kac-Moody superalgebra \cite{Gritsenko:1997au}. Recently, a more physical interpretation of this symmetry was found in the observation that the Weyl group of this superalgebra controls the structure of wall-crossing and discrete attractor flows \cite{Cheng:2008fc}. Furthermore, a physical role of the full algebra was proposed in an equivalence between the dyon degeneracy and a second-quantized multiplicity of a moduli-dependent highest weight of the algebra \cite{Cheng:2008fc}.

Before drawing broader conclusions, it is important to know which of the features mentioned above are of general validity and which are specific to this very special model. To this end we study a larger class of models with $\CN =4$ supersymmetry which are obtained as CHL orbifolds of the toroidally compactified theory. Also in these models, the partition function of BPS dyons is known exactly and is given similarly in terms of Siegel modular forms of congruence subgroups of $Sp(2, \mathbb{Z})$ \cite{Jatkar:2005bh,David:2006ji,Dabholkar:2006xa,David:2006yn,David:2006ru,Dabholkar:2006bj}. The question we would like to address is whether these Siegel forms can be  related to the denominator of some  generalized Borcherds-Kac-Moody superalgebra and  the whether the Weyl group of this algebra can be given a physical interpretation in terms of attractor flows as in the case of the un-orbifolded theory\footnote{This question has  been addressed independently in a recent paper \cite{Govindarajan:2008vi}. However, the results reported here are different in essential ways, particularly regarding the role of the Weyl group.}.

Generalized hyperbolic Kac-Moody algebras of  similar type have made their appearance in string theory in other related contexts. Harvey and Moore have proposed  that a generalized Kac-Moody superalgebra can be associated with BPS states in general, and especially in the context of perturbative half-BPS states
\cite{Harvey:1996gc,Harvey:1995fq}. Another hyperbolic algebra $E_{10}$ has made its appearance as the duality group of one-dimensional theory obtained from the  M-theory compactification  on 10-torus. More generally, the role played by the Weyl groups of hyperbolic Kac-Moody algebras in various gravitational theories in the background close to a spacelike singularity has led to conjectures about the presence of some hyperbolic Kac-Moody symmetries underlying these (super-)gravity theories. See \cite{Gebert:1994zq,Damour:2001sa,Damour:2002cu,Damour:2002et,Henneaux:2007ej,Henneaux:2008me} and references therein\footnote{The Weyl groups which appear in these works are very similar to the Weyl groups of the algebras discussed in the present paper. However, an important difference is that the hyperbolic Kac-Moody algebras discussed there are not Borcherds-Kac-Moody algebras, in that they do not have imaginary simple roots.}, and also \cite{Giveon:1990er,Giveon:1994fu} for earlier discussions of an infinite dimensional gauge algebra underlying the toroidally compactified heterotic string theory.

One difficulty in dealing with the hyperbolic algebras in general is that while their formal structure can be defined in parallel with finite or affine Lie algebras, very often it is not easy to determine all root multiplicities. Without such a knowledge, it is hard to even  begin to find any use of these algebras. As we will see, the situation in the context of $\CN=4$ dyons is better. We will show explicitly that the Weyl denominator identity is satisfied  and provide a complete list of all roots with an algorithm for computing the root multiplicities. We hope that this explicit construction of the algebra, together with the analysis of its role in the BPS partition function and in the physics of crossing the walls of marginal stability, will be a step towards untangling further symmetries in the $\CN=4$ superstring theories.

The paper is organized as follows.
In section $\S${\ref{Summary}}, we summarize the background and our results.  In $\S${\ref{Discrete}} we perform the macroscopic supergravity  analysis of the walls of marginal stability and show that there is a qualitative difference between $\Z_{\SS N}$ models with $N < 4$ and $N \geq 4$.
For $N = 1, 2, 3 $, we determine the Weyl group, the simple real roots, the real root part of the generalized Cartan matrix, Weyl group, and the Weyl chamber of a Borcherds-Kac-Moody superalgebra associated with the dyon spectrum. In $\S${\ref{Exact}} we turn to the microscopic analysis and review the relevant properties of the  exact partition function of BPS dyons for the CHL as Siegel modular forms and write down their expressions as  infinite products and infinite sums.
$\S${\ref{Algebra}} contains some of the main results of the paper, where we show that the automorphic form $\Phi_t$ can be interpreted as the denominator in the Weyl-Kac character formula for the same algebra. In particular, we show that the Weyl denominator identity is satisfied using the sum and the product representations of this automorphic form. This enables us to determine the full root system in principle. After constructing the algebras we  briefly summarize their physical significance.
Finally we conclude with a discussion in $\S${\ref{Discussion}}.

\section{Summary of Results}\label{Summary}

We consider compactifications of the heterotic string to four dimensions on CHL orbifolds
\cite{Chaudhuri:1995fk,Chaudhuri:1995bf,Chaudhuri:1995dj} of the type
\begin{equation}\label{compactification}\frac{ T^5 \times S^1}{\mathbb{Z}_{\SS N}}\, ,
\end{equation}
or equivalently their Type-II duals \cite{Schwarz:1995bj,Vafa:1995gm},
where the generator of the group $\mathbb{Z}_{\SS N}$ acts on the heterotic theory by a $1/N$-shift along the circle $ S^1$ and an order-$N$ left-moving twist symmetry of the $ T^5$-compactified string.
Now, the un-orbifolded toroidally compactified theory has $\CN=4$ supersymmetry and its massless spectrum consists of one graviton multiplet together with $22$ vector multiplets. Since the orbifold symmetry acts trivially on the right-moving fermions, all sixteen supercharges are preserved by the orbifold. On the other hand, since the twist symmetry acts nontrivially on the left-moving gauge degrees of freedom, some of the $22$ vector multiplets will be  projected out. Furthermore, because of the $1/N$ shift, the twisted states have a $1/N$ fractional winding along the
circle and hence all twisted states are massive. As a result, the
orbifolded theory has  fewer
\be
n_v =\frac{48}{N+1}-2
\ee
massless vector multiplets.
The CHL models of interest here are given by $N=1, 2, 3, 5, 7$  and  have $n_v= 22, 16, 12, 8, 4$ vector multiplets, respectively.

The S-duality group of the
$\mathbb{Z}_{\SS N}$ model is the following congruence subgroup $\G_1(N)$ of $PSL(2,\mathbb{Z})$
\begin{equation}\label{Sgamma}
\G_1(N) =\left\{ \bem
       a &b \\
       c & d \eem \Big\lvert
   \quad ad-bc =1,\; \quad c = 0\, \textrm{mod}\, N, \quad
     a = 1\, \textrm{mod}\, N \right\}/\{\pm {\mathds 1}\} \, .
\end{equation}
The  T-duality group of this theory is a subgroup of
\begin{equation}\label{Tdulitygroup}
    O(n_v, 6; \mathbb{Z}).
\end{equation}
A dyonic charge of this theory is specified by a charge vector
\begin{equation}\label{chargevector}
    \bem Q \\ P \eem\,,
\end{equation}
where the components $Q$ and $P$ can be regarded as the electric and magnetic charge vectors of the dyon respectively, each of which transforms in the $(n_v +6)$-component vector representation of the $O(n_v, 6; \Z)$ T-duality group.

The T-moduli live in the Narain moduli space and are parametrized by a vierbein matrix $\m$ which  specifies the right-moving and left-moving projections of lattice vectors like $Q$ and $P$.
For example,
\begin{equation}\label{projections}
    Q_R^a = \mu^a_I Q^I, \quad Q_L^{\til{a}} =  \mu^{\til{a}}_I Q^I, \quad I=1,\ldots, 6+ n_v, \quad a =1, \ldots, 6, \quad {\til{a}}= 1, \ldots, n_v,
\end{equation}
so that $Q^2 = Q_R^2 - Q_L^2$ as usual.
Similarly, the S-moduli live on the upper half-plane and are parametrized by the axion-dilaton field $\lambda$.

The S-duality group (\ref{Sgamma}) acts on the charges and the S-modulus $\l$ as
\begin{equation}\label{charge}
    \left(
                            \begin{array}{c}
                               Q \\
                               P \\
                             \end{array}
                           \right) \rightarrow
                       \left(  \begin{array}{cc}
       a &b \\
       c & d \\
     \end{array} \right)
      \left(
                            \begin{array}{c}
                               Q \\
                               P \\
                             \end{array}
                           \right) \;,\;\; \l \to \frac{a\l+b}{c\l+b}\;\;,\;\;\;\;\bem a & b \\ c & d\eem  \in \G_1(N).
\end{equation}
One can check that restriction on the integers $a$ and $c$ in (\ref{Sgamma}) arises from requirement of preserving the lattice (\ref {quant}) \cite{Vafa:1995gm,Aspinwall:1995fw,Sen:2005iz,Jatkar:2005bh}. We will see later how this S-duality group can be nicely combined with the parity-change transformation into a larger ``extended S-duality group".

The degeneracy of BPS dyons in this class of theories is summarized by a partition function which is the inverse of a Siegel modular form $(\Phi_t)^2$ of a congruence subgroup of $Sp(2, \mathbb{Z})$ of weight $  2  t = \frac{24}{N+1} -2$.
For the cases $N =1, 2, 3$ of most interest for our purposes, the Siegel form has a  square-root  $\Phi_{t}$ which is a Siegel modular form with integral weights $t=5,3,2$.

From the charge vector $\big(\begin{smallmatrix}Q\\P\end{smallmatrix}\big)$, the three quadratic T-duality invariants can be organized as a $2 \times 2$ symmetric matrix\footnote{Apart from the three quadratic invariants listed here, the duality orbits of general $\CN=4$ dyons can depend on additional discrete invariants such as $I =\gcd{ Q \wedge P}$ \cite{Dabholkar:2007vk,Banerjee:2008pv,Banerjee:2008pu, Banerjee:2008ri}. We restrict our attention here to the case of $I=1$.}
\be\label{matrix_charge_vector}
\L_{Q, P}  =\bem Q\cdot Q & Q\cdot P \\
    Q\cdot P & P\cdot P \eem
\;.
\ee
For the twisted states, the T-duality invariants are quantized so that $\L_{Q, P}$ is of the form
\begin{equation}\label{quant}
  \left(
  \begin{array}{cc}
    2n/N & \ell \\
    \ell & 2m \\
  \end{array}
\right),
\end{equation}
with $n, m, \ell$ all integers.

Recall that the S-duality groups (\ref{Sgamma}) are subgroups of $PSL(2,\Z)$, which is the same as  the time-orientation preserving component $SO^+(2, 1; \Z)$  of the Lorentz group acting on  a Lorentzian space $\R^{2,1}$. Since the charge vector $\big(\begin{smallmatrix}Q \\ P\end{smallmatrix}\big)$ transforms as a doublet  in the spinor representation of $SO(2,1)$, the vector $\Lambda_{Q,P}$ of T-duality invariants  transforms as a triplet in the vector representation of this Lorentz group.
Therefore, the matrices  (\ref{matrix_charge_vector}) with the quantization (\ref{quant}) form a lattice
in a Lorentzian space $\R^{2,1}$.
One of our main results is to show that, in some cases,  this Lorentzian lattice spanned by all possible T-duality invariants $\L_{Q,P}$  can be interpreted as the root lattice of a Borcherds-Kac-Moody superalgebra. In this respect, the models with $ N<4$ and $N\geq 4$ are qualitatively different both physically and mathematically,  as will be explained in detail in $\S${\ref{Discrete}}. More specifically, only in the models with $ N<4$, the walls of marginal stability for the dyons partition the moduli space into compartments bounded by finitely many walls.

\subsection{Models with $ N < 4$}

From the {macroscopic} supergravity analysis it is known that for the $N<4$ models,  the walls divide the moduli space into connected domains each bounded by a finite number of walls for a given set of total charges\cite{Sen:2007vb}. We will combine  these results with a {microscopic} analysis to identify a Borcherds-Kac-Moody superalgebra with the following features.
\begin{myitemize}
\item In $\S${\ref{Discrete}}, we identify a special set of lattice vectors $\{ \alpha_i^{\scriptscriptstyle{(N)}}\}$ with $i = 1, \ldots r^{\scriptscriptstyle{(N)}}$ where  $r^{\scriptscriptstyle{(N)}} = 3, 4, 6$ for $N = 1, 2, 3$ respectively. These lattice vectors will eventually be identified as the real simple roots of the relevant Borcherds-Kac-Moody algebra.
\item
For each model, the matrix of inner products of the simple real roots defines a symmetric Cartan matrix
\begin{equation}\label{cartan}
   A^{\scriptscriptstyle{(N)}}_{ij} = (\a_i^{\scriptscriptstyle{(N)}},\a_j^{\scriptscriptstyle{(N)}}); \quad i, j = 1, \ldots r^{\scriptscriptstyle{(N)}} \, .
\end{equation}
The Cartan matrices in the three cases are  given by
      \be\label{cartan1}
A^{(1)} = \left( \begin{array}{rrr}2&-2&-2\\ -2&2&-2\\ -2&-2&2\end{array}\right),
\ee
  \be\label{cartan2}
A^{(2)} = \left(
\begin{array}{rrrr} 2&-2&-6&-2\\
-2&2&-2&-6\\
-6&-2&2&-2\\
-2&-6&-2&2
\end{array}\right) ,
\ee
\be\label{cartan3}
A^{(3)} =\left(
\begin{array}{rrrrrr}
2&-2&-10&-14&-10&-2\\
-2&2&-2&-10&-14&-10\\
-10&-2&2&-2&-10&-14\\
-14&-10&-2&2&-2&-10\\
-10&-14&-10&-2&2&-2\\
-2&-10&-14&-10&-2&2
\end{array}
\right),
\ee
respectively in the three cases $N=1, 2, 3$.
\item
All three  Cartan matrices are `hyperbolic' in that they have one negative eigenvalue. All  have rank three. Now, rank three, hyperbolic Cartan matrices have been  classified  by  Gritsenko and Nikulin \cite{Gritsenko:1996-1,Gritsenko:1996-2} with two additional `niceness' assumptions:

  (a) that they are `elliptic' in that the fundamental Weyl chamber has  finite volume,

  (b) that they admit a lattice  Weyl vector defined by (\ref{def_weyl_vector}).

The set of such hyperbolic Cartan matrices is a finite list and the three Cartan matrices $A^{(1)}, A^{(2)}, A^{(3)}$ above
correspond to $A_{II,1}, A_{II,2}, A_{II,3}$  in this classification. Moreover, they are unique in that  these are the only Cartan matrices in the classification with all the vertices of the Weyl chamber at infinity. In our context, the condition of having all vertices at infinity follows from the physical consideration that the walls of marginal stability of the theories under consideration have all intersections at the boundary of the moduli space and they do not terminate in the middle of the moduli space.
%As we will see, the requirement of the existence a lattice Weyl vector %is crucial for the automorphic properties of the partition function %\cite{{Borcherds:1998au},Harvey:1995fq}.

\item
The group of reflections with respect to the simple real roots defines the Weyl group $W^{\scriptscriptstyle{(N)}}$ of the model for each $N$.
As mentioned earlier, for a given set of  charges, the moduli space of these models is divided into connected domains each bounded by a finite number of walls.  All domains can be mapped to the `fundamental Weyl chamber' by the action of this Weyl group. Physically, the `fundamental Weyl chamber' is the attractor region in the moduli space where no multi-centered solution which can decay exists. From the number of simple real roots $r^{\SS (N)}=3,4,6$, we see that the fundamental Weyl chamber mapped onto the Poincaré disk is  triangular, square, and hexagonal for $N=1, 2, 3$ respectively (Fig. \ref{Walls}).

\item
The Weyl group, generated by the reflections
\be\label{generator_weyl}
s_i^{\scriptscriptstyle{(N)}} : X \to X-  2\a_i^{\scriptscriptstyle{(N)}} \frac{(X,\a_i^{\scriptscriptstyle{(N)}})}{(\a_i^{\scriptscriptstyle{(N)}},\a_i^{\scriptscriptstyle{(N)}})}\;\;,\;\; i = 1,\dotsi,r^{\SS (N)}\;,
\ee
can be directly related to the physical symmetry group of the theory as follows. The extended physical symmetry group of each $\mathbb{Z}_N$ model is  the following congruence subgroup of $PGL(2,\Z)$
 \be\label{tildeduality}
\til\G_1(N) = \bigg\{\bem a & b \\ c& d\eem \Big\lvert  \,
ad-bc = \pm 1, c= 0 \text{ mod }N\;, a= \pm 1 \text{ mod }N
\bigg\}/\big\{ \pm {\mathds   {1}}\big\} \;
\ee
which is generated by the $\Z_2$-symmetry of parity change and the S-duality group. This physical symmetry group turns out to be also the symmetry group of the root system of the relevant Borcherds-Kac-Moody algebra. More precisely, it is related to the Weyl group as
\be\label{S_Weyl_group}
\til\G_1(N) = W^{\scriptscriptstyle{(N)}} \rtimes \text{\small Sym}({\cal W}^{\scriptscriptstyle{(N)}})\;\;,\;\;N=1,2,3\;,
\ee
where   $ \text{\small Sym}({\cal W}^{\scriptscriptstyle{(N)}})$ is a finite group of symmetries of the corresponding fundamental Weyl chamber preserving the relevant lattice structure (\ref{quant}).

\item
We show that for each theory there exists a Weyl vector satisfying (\ref{def_weyl_vector}).
On the other hand, the dyon degeneracies of $N=1,2,3$ models are given in terms of a genus two Siegel modular form $\Phi_{t}(\O)$ as in (\ref{PF0}).
With this Weyl vector,  the Siegel modular form $\Phi_{t}$ can be interpreted as the denominator of  a character of a representation of a Borcherds-Kac-Moody superalgebra, and the Weyl vector is related to the level-matching condition of the heterotic string. Moreover, the Weyl denominator identity is manifestly satisfied by the Siegel modular form $\Phi_{t}$, as  can be seen from
its  product and sum representation. Some roots are bosonic and others are fermionic and they lead to a large cancelation in the index.  This identity provides an algorithm for determining  the multiplicities of both imaginary and real roots of this BKM superalgebra. In $\S${\ref{Algebra}} we explicitly determine the root multiplicities of some low-lying roots.
\item
Based on this analysis, we propose a microscopic model for the quarter-BPS dyons for all three cases along the lines of  \cite{Cheng:2008fc}. The Weyl group of this algebra is identified with the group of attractor flows. The dyon degeneracy is given by the ``second-quantized multiplicity" of a charge- and moduli-dependent highest weight vector.
Specifically, apart from being consistent with the expected asymptotic growth of degeneracies in the large charge regime, such a proposal predicts a jump in the (graded) dyon degeneracy when the moduli cross a wall of marginal stability that  precisely reproduces the result from macroscopic analysis.
\end{myitemize}

\subsection{Models with $N \geq 4$}

For the $\Z_{\scriptscriptstyle N}$ models with $N \geq 4$, the situation is much less clear. {}From the macroscopic supergravity analysis, we conclude that the walls are so severely removed due to the orbifold action such that the partition of the moduli space by the walls of marginal stability yields compartments which are not bounded by finitely many walls. On the other hand, on the microscopic side, partition functions for quarter-BPS dyons have been proposed for prime numbers $N=5,7$ in terms of Siegel modular forms \cite{Jatkar:2005bh}. From our analysis, it appears impossible to relate these partition functions to the denominators of any generalized Kac-Moody algebra.

It is not clear to us how to interpret these results for $N \geq 4$. One possibility is that in these cases, the symmetry is of an even  more general type. If this is the case, then the symmetry might be too general to be particularly illuminating. A more radical possibility is that these models are non-perturbatively inconsistent  even though they are perfectly sensible as perturbative orbifolds.

In any case, the $N<4$ models provide a class of models for which the BPS dyon spectrum exhibits  an immense symmetry. To construct and interpret the Borcherds-Kac-Moody algebra for this class of models will be the focus of the present paper.

\section{Discrete Attractor Flows and Weyl Groups} \label{Discrete}

In this section we first quickly review the analysis of the relationship between the real roots of the algebra and the walls of marginal stability in the un-orbifolded theory \cite{Cheng:2008fc}.
After introducing the setup, we redo the supergravity analysis \cite{Sen:2007vb} of the walls of marginal stability for the CHL models from a more group-theoretic and algebraic point of view.  In particular we study how the walls divide the future light-cone, and recover the result that the moduli space is divided into adjoining domains each bounded by a finite number of walls if and only if $N <4$. From the structure of the walls we can then define the group of wall-crossings, generated by the reflections with respect to the set of walls bounding one domain. As was argued in \cite{Cheng:2008fc}, the attractor flow of the moduli fields in a black hole solution with the given charges has to follow the ordering of such a group, and by construction the full moduli dependence of the BPS spectrum is encoded in this ``group of discrete attractor flow".

\subsection{The Walls of Marginal Stability}

To make the Lorentz group action of the duality group $PSL(2,\Z) \cong SO^+(2,1,\Z)$ of the toroidally compactified heterotic theory more manifest, we identify the Lorentzian space $\R^{2,1}$ with the space of \(2 \times 2\) symmetric matrices $X$ with real entries equipped with the metric
\be\label{metric_M2_space}
(X,X)= -2\, \text{\small det} X\;,
\ee
and we choose the time-orientation  such that the future light-cone is given by
\be
V^+ = \left\{X\Big\lvert \;X=\bigg( \begin{array}{ll} x^+ & x\\ x & x^- \end{array}\bigg),\; X=X^T\;, \Tr X > 0 \;, \text{\small det}X > 0\right\}\;.
\ee
Note that with $x^{\pm} = t\pm y$, the metric
\begin{equation}\label{lormetric}
   (X, X) = -2 \det X  =  2(-t^2 + x^2 + y^2)
\end{equation}
is indeed the usual Lorentzian metric in $\mathbb{R}^{2,1}$. The normalization factor of $2$ is chosen consistent with the group theory convention that simple roots have  length-squared two. For later notational convenience we also define $ |X|^2= -(X,X)/2 =\text{\small det} X  $.

In this space, a discrete Lorentz transformation is given by
\be\label{action_PGL2Z}
X \to \g(X) := \g X \g^T
\ee
for some $PSL(2,\Z)$ matrix $\g$.
The vector of T-duality invariants is given by the symmetric $2 \times 2$ matrix $\Lambda_{Q, P}$ (\ref{matrix_charge_vector}). When a charge vector transforms as
\be \bem Q \\ P \eem \rightarrow \gamma \bem Q \\ P \eem \;,\ee
the matrix of T-duality invariants transforms as $$\Lambda_{Q, P} \rightarrow \gamma (\Lambda_{Q, P})\, .$$
The length of this vector is related to the Bekenstein-Hawking entropy
\be
S(P,Q) = \p |\L_{Q,P}|  = \p \sqrt{Q^2 P^2 - (Q\cdot P)^2}
\ee
of the black hole with charge $(Q,P)$. Note that  half-BPS states correspond to vectors on the boundary the future light-cone whereas quarter-BPS states correspond to vectors in its interior.

Besides the charge vector $\L_{Q,P}$, there are other  vectors that live naturally in the future light-cone. First, one can define the right-moving charge vector
\be\label{rightmatrix_charge_vector}
\L_{Q_R, P_R}  =\bem Q_R\cdot Q_R & Q_R\cdot P_R \\
    Q_R\cdot P_R & P_R\cdot P_R \eem
\;.
\ee
which implicitly depends on the T-moduli $\mu$ through (\ref{projections}). One can then define  two natural vectors associated to the axion-dilaton S-moduli $\lambda$ and the T-moduli $\mu$ respectively by
\be\label{ST_moduli}
{\cal S} = \frac{1}{\Im\l} \bem|\l|^2 & \Re\l \\\Re\l & 1 \eem\quad,\quad {\cal T} =  \frac{1}{{|\L_{Q_R, P_R}|}} \L_{Q_R, P_R}\;.
\ee
Both matrices are normalized to unity $|X|^2=1$ and transform as $X \to \g(X)$ under S-duality transformation (\ref{extended_S-dual}).
In terms of them, we can construct the moduli-dependent `central charge vector'
\be\label{def_Z_vec}
{\mathcal Z} = {\scriptsize\sqrt{|\L_{Q_R, P_R}|}}\, \big({\cal S}+{\cal T}\big) ,
\ee
whose norm equals the BPS mass
\be
M_{Q,P} = |{\mathcal Z}|
\ee
and whose orientation satisfies
$
{\mathcal Z}\lvert_{\text{attr.}} \sim\L_{Q,P}
$
at the attractor moduli of the black hole of  charge $(Q,P)$.

Because the central charge vector enters the BPS mass formula, it is probably not surprising that it encodes  all stability information of all potential multi-centered configurations with the total charge $(Q,P)$ which are relevant for the counting. One can show that these are the two-centered solutions with each center carrying $1/2$-BPS charges \cite{Sen:2007nz} in the following form \cite{Sen:2007vb,Cheng:2008fc}
\bea\notag \label{split_1}
\bem Q \\ P\eem& =&\bem Q_1 \\ P_1\eem +\bem Q_2 \\ P_2\eem \\&=& \pm  \left\{ (aQ-bP) \bem d \\ c\eem
+ (-cQ+dP) \bem b \\ a\eem \right\}\;\;,\;a,b,c,d \in \Z,ad-bc = \pm1\;.
\eea
As promised, the stability condition on the moduli for the corresponding two-centered solution to exist can be expressed solely in terms of the moduli vector ${\mathcal Z}$:
\be\label{stab}
(\a,{\mathcal Z}) (\a, \L_{Q,P}) < 0
\ee
with
\be\label{root_split_1}
\a = \bem 2bd & ad+bc \\ ad+bc & 2ac\eem\;.
\ee
The boundary of the above stability region is given by the marginal stability condition
$M_{Q,P}= M_{Q_1,P_1}+M_{Q_2,P_2}$, which can be rewritten as
\be\label{WMS}
({\cal Z},\a) =0
\ee
and which defines a surface of codimension one in the moduli space called the ``wall of marginal stability" for the two-centered solution with charges $(Q_1,P_1)$ and $(Q_2,P_2)$\footnote{See also \cite{Denef:2000nb,Denef:2002ru,Denef:2007vg,Gaiotto:2008cd}  for studies of the wall of marginal stability in a $\CN=2$ setting.}.
In this form it is manifest that none of these two-centered solutions exist when the moduli are at their attractor values, since  ${\mathcal Z}\sim \L_{Q,P}$ and equation (\ref{stab}) cannot be satisfied.

It was observed in \cite{Cheng:2008fc} that the above vectors $\a$ can be identified with the positive real roots of the underlying Borcherds-Kac-Moody algebra. Recall that there are different choices of positive roots, or equivalently different choices of simple roots in the root system, which one can make in a Lie algebra and which are related to one another by a Weyl reflection. For counting dyons, the choice is fixed in terms of the total charges by the condition
\be
(\L_{Q,P},\a) < 0\;\;\text{for all positive roots  }\a \in \D_+\;,
\ee
or equivalently that the charge vector $\L_{Q,P}$ lies inside the ``fundamental Weyl chamber" \footnote{Note that this is actually a condition on the choice of {\it real} simple roots only, since the condition is automatically satisfied as long as both vectors $\L_{Q,P}$ and $\a$ are in the future light-cone $V^+$. Unless otherwise noted, we will restrict our attention to charges with a classical horizon ($\L_{Q,P} \in V^+$) and also exclude those charges which admit two-centered scaling solutions ($(\L_{Q,P},\a) =0$ for some real root $\a$)
in this paper. }
\be\label{funda_chamber_1}
{\cal W}= \left\{X\Big\lvert \;(X,\a) \leq 0  \text{    for all positive roots    }
\a \right\}\;.
\ee

In a more abstract notation, the
positive real roots $\a$ of the algebra  are in one-to-one correspondence with the two-centered solutions with the charges given by
\bea\nonumber
\L_{Q_1,P_1} &=& P_\a^2\,\a^+ \quad,\quad\L_{Q_2,P_2} =  Q_\a^2\,\a^-\\
\label{charge_split_intro}
\L_{Q,P} &=& P_\a^2\,\a^+ + Q_\a^2\,\a^-  - |(P\cdot Q)_\a|\,\a\;,
\eea
where the lightlike vectors
\be\label{alpha_pm}
\{\a^+,\a^-\} =
 \Bigg\{ \bem b^2 & ab \\ ab & a^2\eem, \bem d^2 & cd \\ cd & c^2\eem \Bigg\}
\ee
are determined as the vectors on the intersection of the light-cone with the plane of marginal stability (\ref{WMS})
and normalized to have inner product $(\a^+,\a^-)=-1$.

To relate the walls of marginal stability in the moduli space to the objects of the BKM algebra, recall that the Weyl group, which we will call $W$ (or $W^{\SS (N)}$ for general $N>1$), is defined to be the group of reflections with respect to the real roots and divides the Lorentzian root space into different Weyl chambers. In particular, the fundamental Weyl chamber ${\cal W}$  is defined as in
(\ref{funda_chamber_1}). In our case, when we identify the Minkowski space $\R^{2,1}$ with the projected moduli space of the moduli vector ${\cal Z}$, the above statement means that the walls of the Weyl chambers are exactly the physical walls of marginal stability and crossing a wall can be described as a Weyl reflection with respect to the positive root corresponding to the two-centered configuration in question. Specifically, the fundamental Weyl chamber ${\cal W}$ is the so-called ``attractor region" where no relevant two-centered configuration exists, and an attractor flow can be identified with a series of Weyl reflections from elsewhere in the future light-cone to ${\cal W}$.

Another physical way to understand the appearance of the Weyl group lies in its relation to the physical duality group. The walls of marginal stability cuts the future light-cone into different domains bounded by sets of, in the present un-orbifolded case, three walls. Put differently, when projecting the future light-cone onto a constant-length hyperboloid, the Weyl group gives a triangular tessellation of the Poincar\'e disk in which the Weyl chamber ${\cal W}$ is the triangle bounded by three planes of orthogonality to the three simple real roots of the algebra, whose matrix of inner products (\ref{cartan1})
is the real root part of Cartan matrix of the Borcherds-Kac-Moody algebra. See Figure \ref{Walls}.

 Then we find that the full symmetry group of the root system is
$PGL(2,\Z) = W \rtimes \text{\small Sym}({\cal W})\;,
$
 where $\text{\small Sym}({\cal W})$, the symmetry group of the fundamental domain which leaves the appropriate lattice structure intact, is in this case the symmetry group $D_3$ of a regular triangle. On the other hand, the group
 \be\label{pgl2z} PGL(2,\Z)=\left\{ \bem
       a &b \\
       c & d \eem \Big\lvert
   \;\; ad-bc =\pm 1 \right\}/\big\{ \pm {\mathds   {1}}\big\}\ee
 is nothing but the duality group $PSL(2,\Z)$ discussed before, now extended with the spacetime parity change operation generated by
 \be\label{parity_change_ope}
 R: \,\bem Q \\ P \eem \to  \bem Q \\ -P \eem \;\;
,\;\; \l \to -\bar{\l}
 \ee
 and acts on the charges and the heterotic axion-dilaton as
\begin{align}\nonumber
&\bem Q\\P\eem \to \bem a&b\\c&d \eem \bem Q\\P\eem\\[4mm]
\label{extended_S-dual}
&\l \to  \left\lbrace\begin{array}{ll}\displaystyle \frac{a\l+b}{c\l+d} \; & \;\;\text{when}\;\; ad-bc=1\\[4mm]  \displaystyle \frac{a\bar{\l}+b}{c\bar{\l}+d}\; & \;\;\text{when}\;\; ad-bc=-1  \;\;\end{array} \right. \ .
\end{align}

This concludes our review of the role of the positive real roots, the Weyl group, and the Weyl chamber of a Borcherds-Kac-Moody algebra in the un-orbifolded theory.

\subsection{Three Arguments for Two Classes of Theories}
\label{Three Arguments for Two Classes of Theories}

For the $N=1$ case discussed in \cite{Cheng:2008fc}, the set of three simple real roots which gives the Cartan matrix (\ref{cartan1}) can be chosen to be
\be\label{simple_real_root_1}
\a_1 = \bem 0 & -1 \\ -1 & 0\eem,\;\; \a_2 =\bem 2 & 1 \\1 & 0\eem,\;\;
\a_3 =\bem 0 & 1 \\1 & 2\eem\;.
\ee
The corresponding positive real roots take the form
\be
\a = \bem 2n& \ell \\ \ell & 2m \eem, \;(\a,\a) =  2\;\;,\;\;(n,m,\ell) > 0
\ee
where the condition $(n,m,\ell)> 0$ on the integers means $n,m \geq 0$, $\ell\in \Z$ and $\ell <0$ when $n=m=0$. From the condition that the charge vector $\L_{Q,P}$ has to lie in the fundamental Weyl chamber (\ref{funda_chamber_1}), this choice of the simple roots is actually only suitable for dyon charges whose T-duality invariants satisfy $Q^2,P^2 > Q\cdot P > 0$. But from the relation between the Weyl group and the physical duality group (\ref{S_Weyl_group}), we see that it is always possible to map the charges such that the above set of simple roots is the right choice for the charges.

When we take the orbifold, because the T-duality invariants are now quantized as (\ref{quant}) due to the presence of the twisted states, not all the splits of charges in (\ref{split_1}) are allowed for forming a two-centered solution. Instead we have to restrict ourselves to elements of the following congruence subgroup of $PGL(2,\Z)$
\be\label{tilg0}
\til\G_0(N) = \bigg\{\bem a & b \\ c& d\eem \Big\lvert  \,
ad-bc = \pm 1, c= 0 \text{ mod }N\;
\bigg\}/\big\{ \pm {\mathds   {1}}\big\} \;.
\ee

Notice that there is an extra constraint on the group element for the extended S-duality group $\til\G_1(N)$ since it has to leave invariant the spacing
of the lattice (\ref{quant}) of all possible vectors $\L_{Q,P}$ of T-duality invariants. Only for $N<4$ these two groups $\til\G_0(N) $ and $\til\G_1(N) $
happen to be the same.

From the relation between the split of charges and the root (\ref{root_split_1}), the charge quantization imposes that the roots relevant for the wall-crossing must be of the form
\be\label{condition_alpha}
\a^{\scriptscriptstyle{(N)}} = \bem 2* & * \\ * & 2N* \eem\;\;,\;\;(\a,\a)=2\;.
\ee
Therefore, we expect that the ``positive real roots" of the
$\Z_{\SS N}$-orbifolded theory will be of the form\footnote{To avoid proliferation of notations and names, we are already using in our macroscopic analysis the terminology for the microscopic algebra which will be defined in later sections.}
\be\label{posi_root_N}
\a^{\scriptscriptstyle{(N)}} = \bem 2n& \ell \\ \ell & 2 m \eem, \;(\a,\a) =  2\;,\;\;(n,m,\ell) > 0\;, m =0 \text{  mod   }N\;\;\;.
\ee

From this we immediately see that, of the three simple roots of the un-orbifolded theory (\ref{simple_real_root_1}), two of them $\a_{1,2}$ always survive the orbifolding while the third one gets removed whenever $N>1$. In other words, if we denote $\{\a_i^{\scriptscriptstyle{(N)}}\}$ the set of ``simple real roots" whose walls of orthogonality bound a domain in the moduli space, then we have
\be
\a_1^{\scriptscriptstyle{(N)}}= \a_1 \;\;,\;\;\a_2^{\scriptscriptstyle{(N)}}= \a_2 \;\;\text{   for all  } N \;.
\ee
A particular symmetry of the positive real roots will help us to find the rest of the simple roots corresponding to the other walls which bound a domain in the moduli space together with the two walls $(\a_{1,2},{\cal Z})=0$ found above. Notice that there is a special element of the $\til\G_1(N)$ group
\be\label{sym_weyl_chamber}
\g^{\scriptscriptstyle{(N)}} = \bem 1 & -1 \\ N & 1-N \eem \;,
\ee
which has the property that when it acts on a positive root of the form  (\ref{posi_root_N}), the image is again a positive root satisfying the conditions in (\ref{posi_root_N}). This means that this transformation permutes simple roots among themselves and must therefore be a symmetry of the fundamental Weyl chamber ${\cal W}^{{\scriptscriptstyle(N)}}$. For example, for $N=1$ one can easily check that $\g^{(1)}$ acts as a permutation of the three simple roots $\a_{1},\a_{2},\a_{3}$ and is indeed a symmetry of the simple root system.

Now we are ready to show how the existence of such a symmetry predicts that the walls of marginal stability fail to partition the moduli space into finite compartments when $N > 3$. Concretely, the existence of this symmetry implies that
\be
\g^{{\scriptscriptstyle(N)}}(\a_1), \g^{{\scriptscriptstyle(N)}}(\a_2) \in \{\a^{{\scriptscriptstyle(N)}}_i\} \;,
\ee
so a necessary condition for the set of walls bounding the fundamental
Weyl chamber to be finite is that $(\g^{{\scriptscriptstyle(N)}})^k=\mathds{1}$ for some finite integer $k$. There are many ways to see that it is not the case for $N>3$. We will sketch three ways to understand this which give us different insights into the topology of the walls of marginal stability of the CHL theories.

%\vskip .4cm
\subsubsection{\it A Group Theoretic Argument}
%\vskip .2cm

Recall that the group $PSL(2,\Z)$ is generated by the two generators $S$ and $ ST$, and the only relations among them are
\be
S^2 = (ST)^3 =  {\mathds   {1}}\;.
\ee
For convenience let us write
$
B= ST
$.
Now observe that
\be
\g^{{\scriptscriptstyle(N)}} = (SB^2)^{N-1} SBS = (SB^2)^{N-1} S(SB^2)^{-1}
\ee
and thus
\bea
(\g^{{\scriptscriptstyle(N)}})^k &=& (SB^2)^{N-1} [ S (SB^2)^{N-2} ]^{k-1} S(SB^2)^{-1}  \\
&=& (SB^2)^{N-1} [ B^2 (SB^2)^{N-3} ]^{k-1} S(SB^2)^{-1}  \;,
\eea
from this we immediately see that
\be
(\g^{{\scriptscriptstyle(2)}})^2 = (\g^{{\scriptscriptstyle(3)}})^3  =  {\mathds   {1}}\;,
\ee
but there is no $k < \inf$ such that $(\g^{{\scriptscriptstyle(N)}})^k  =  {\mathds   {1}} $ for any $N \geq 4$\;.

\subsubsection{\it A Geometric Argument}

%\vskip .2cm
Recall that  the matrix $\g=\big(\begin{smallmatrix}a&b\\c&d\end{smallmatrix}\big)$ defines a M\"obius transformation on the Riemann sphere
\be\label{mobius}
z \to \frac{az+b}{cz +d}\;.
\ee
By inspecting the eigenvalues of $\g^{{\scriptscriptstyle(N)}}$
\be
\g^{{\scriptscriptstyle(N)}} \xrightarrow[\substack{similarity\\ transf.}]{} \bem \z&0 \\ 0&\z^{-1}\eem\;,
\ee
 it is easy to see that the corresponding M\"obius transformation is elliptic, parabolic, or hyperbolic when
\be
(\Tr \g^{{\scriptscriptstyle(N)}})^2 = (N-2)^2 \;\;
\begin{cases} <4 \\ = 4 \\ > 4
\end{cases}\;.
\ee

In other words, when viewed as a M\"obius transformation of the Riemann sphere, $\g^{\SS (N)}$ is conjugate to
\be
z \to \begin{cases}
e^{i\th}z\\
z+ \th \\
e^{\th}z
\end{cases}\;\;\;
\text{when} \;\;\;  N \;\;\begin{cases} <4 \\ = 4 \\ > 4
\end{cases}\; \text{   for some   } \th \in \R\;,
\ee
and clearly does not return to itself whenever $N \geq 4$.

\subsubsection{\it An Arithmetic Argument}

In the un-orbifolded theory, the relevant two-centered solutions (\ref{split_1}) are given by positive roots of the form (\ref{root_split_1}), or equivalently the lightlike vectors $\a^{\pm}$ given in (\ref{alpha_pm}). Therefore, with each wall of marginal stability one can uniquely associate a pair of rational numbers of the following form $\{\textstyle{\frac{b}{a},\frac{d}{c}}\}$, with the normalization fixed by the condition that $a,b,c,d$ being integers from which $a$ and $c$ are nonnegative and satisfying  $ad-bc=1$\cite{Cheng:2008fc}.

\begin{figure}
\centering
\includegraphics[width=4.9in,height=5.15in]{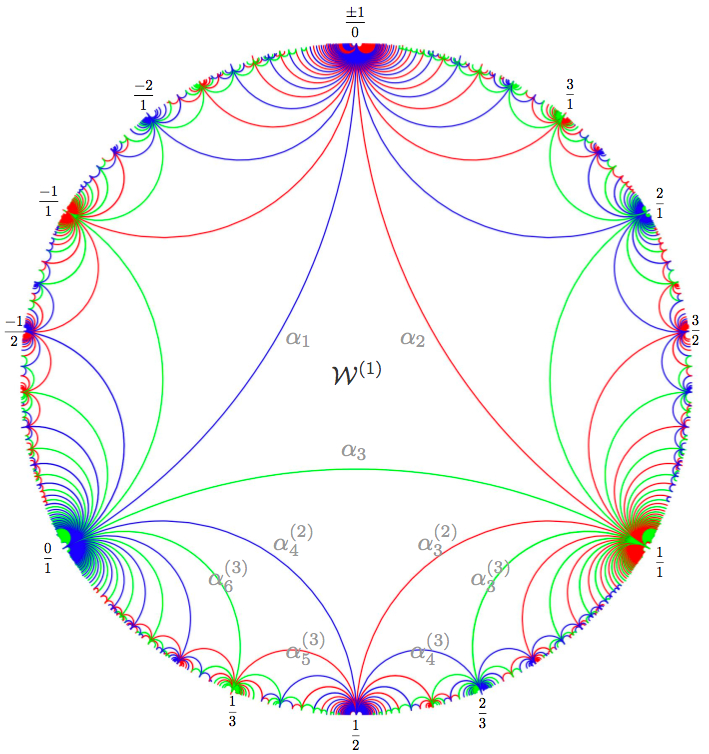}
\caption{The structure of walls when projected onto the Poincaré disk. The fundamental Weyl Chamber is bounded by three walls  labeled by $\alpha_1, \alpha_2, \alpha_3$ for $N=1$; by four walls labeled by $\alpha_1, \alpha_2, \alpha^{(2)}_3, \alpha^{(2)}_4$ for $N=2$; and by six walls labeled by $\alpha_1, \alpha_2, \alpha^{(3)}_3, \alpha^{(3)}_4, \alpha^{(3)}_5, \alpha^{(3)}_6$ for $N=3$. Recall that each triangle is really equivalent and any triangle can be mapped to the central one by some conformal transformation.} \label{Walls}
\end{figure}

This corresponds to compactifying the real line by identifying $+\inf$ with $-\inf$ into a circle, which is then identified with the boundary of the Poincar\'e disk, or equivalently the boundary of the future light-cone $V^+$ before projecting it onto a constant-length slice.
Then the walls of marginal stability of the $N=1$ theory are in one-to-one correspondence with the geodesic lines connecting two neighboring rational numbers in the so-called Stern-Brocot tree
\be
\begin{array}{ccccccccccccc}
\vdots&\vdots& \vdots& \vdots& \vdots& \vdots& \vdots& \vdots& \vdots& \vdots& \vdots& \vdots& \vdots \\
\frac{-1}{0}&\frac{-2}{1}&\frac{-1}{1}&\frac{-1}{2}&\frac{0}{1}&\frac{1}{3}&\frac{1}{2}&\frac{2}{3}&\frac{1}{1}&\frac{3}{2}&\frac{2}{1}&\frac{3}{1}&\frac{1}{0}\\&&&&&&&&&&&& \\
\frac{-1}{0}&&\frac{-1}{1}&&\frac{0}{1}&&\frac{1}{2}&&\frac{1}{1}&&\frac{2}{1}&&\frac{1}{0}\\
&&&&&&&&&&&& \\
\frac{-1}{0}&&&&\frac{0}{1}&&&&\frac{1}{1}&&&&\frac{1}{0}
\end{array}\ee
 which is formed by successively taking the ``mediant"
$
 \frac{b+d}{a+c}
$
of the previous pair of rationals  $\{\textstyle{\frac{b}{a},\frac{d}{c}}\}$ starting from $\frac{\pm1}{0}$, $\frac{0}{1}$, and which contains all the rational numbers. For example, the fundamental Weyl chamber in the un-orbifolded theory is bounded by walls connecting the three rational numbers $\frac{\pm1}{0}$, $\frac{0}{1}, \frac{1}{1}$. See Figure \ref{Walls}.

From this point of view of the rational numbers, the effect of orbifold is that some lines connecting certain pairs of the rationals will be removed by the orbifold and the fundamental Weyl chamber will correspond to a finer spacing of the real line by the rational numbers, such that the walls correspond to the lines connecting two pairs of rational numbers $\{\textstyle{\frac{b}{a},\frac{d}{c}}\}$ with the property that the product of the two denominators are divisible by $N$. Namely, the walls of the $\Z_{\SS N}$ theory will correspond to lines connecting pairs of rational numbers of the form
\be\label{rational_number_1}
\Big\{\frac{b}{a},\frac{d}{c}\Big\}\quad,\quad ad-bc=1\quad,\quad a,c \geq 0,\; ac = 0 \text{   mod   }N\;.
\ee
 In particular, the fact that $\a_1$, $\a_2$ are left untouched while $\a_3$, corresponding to the pair $\{\textstyle{\frac{0}{1},\frac{1}{1}}\}$, is removed for all $N>1$, is translated to the statement that the fundamental Weyl chamber will correspond to a finer spacing of the segment $[0,1] \in \R$  of the real line. Therefore we will now focus on
the following middle part of the Stern-Brocot tree starting from 0 and ending at ${1}$
\be\label{middle_tree}
\begin{array}{ccccc}
\vdots&\vdots& \vdots& \vdots& \vdots\\
\frac{0}{1}&\frac{1}{3}&\frac{1}{2}&\frac{2}{3}&\frac{1}{1}\\
&&&&\\
\frac{0}{1}&&\frac{1}{2}&&\frac{1}{1}\\
&&&&\\
\frac{0}{1}&&&&\frac{1}{1}
\end{array}\;.
\ee
From the symmetry consideration, there is another thing we can say about the division of the segment $[0,1] \in \R$ by the walls of a fundamental Weyl chamber ${\cal W}^{{\scriptscriptstyle(N)}} $. That is, we expect the fundamental Weyl chamber to be a regular polygon, or to say that the Gram matrix $A^{\SS (N)}$ of inner products of the simple roots should be invariant under a cyclic permutation of them. This will correspond to a row of in the above tree (\ref{middle_tree}). For example, as we will see later, the second row $\{\frac{0}{1},\frac{1}{2},\frac{1}{1}\}$ and the third row  $\{\frac{0}{1},\frac{1}{3},\frac{1}{2},\frac{2}{3},\frac{1}{1}\}$ will give us the real simple roots of the $N=2$ and $N=3$ model respectively.
But there is no such a row (\ref{rational_number_1}) with finitely many  elements such that the neighboring rational numbers all have the product of their denominators divisible by $N$, when $N\geq 4$. This excludes the existence of a domain in the moduli space bounded by a finite number of walls in these theories\footnote{According to our macroscopic analysis, for the $N=4$ model the walls of marginal stability divide the moduli space into domains bounded by an infinite number of walls which are related to one another by some S-duality transformation. Partition function for the dyons in such a model has been proposed in \cite{David:2006ud} but the analysis is complicated by the fact that $N$ is not a prime in this case and we will not discuss it in the present paper.}.

\subsection{The Finite Cases}
\label{The Finite Cases}
We now focus on the  $N<4$ models in which the walls of marginal stability partition the moduli space into compartments each bounded by a finite number of walls. We would like to identify the real simple roots, the Weyl group and the fundamental Weyl chamber of the relevant algebra which will be shown to generate the BPS dyon spectrum of the theory.

\subsubsection{$N=1$}

This case was analyzed in \cite{Cheng:2008fc}, where it was  found that three simple roots given by (\ref{simple_real_root_1}) and the physical extended S-duality group is also the symmetry group of the root system, which is related to the Weyl group as (\ref{S_Weyl_group}).

\subsubsection{$N=2$}

As mentioned earlier,
the simplest way to read off the real simple roots for a given $N$ is to read them off from the relevant part of the Stern-Brocot tree of rational numbers. From (\ref{root_split_1}) and the second row in (\ref{middle_tree}) we see that,
apart from $\a^{\scriptscriptstyle (2)}_1=\a_1$ and $\a^{\scriptscriptstyle (2)}_2=\a_2$ as in the un-orbifolded case in (\ref{simple_real_root_1}), there are two other real simple roots
\be\a_3^{(2)} = \bem 2 &3 \\ 3&4\eem \;\;,\;\;
\a_4^{(2)} = \bem 0 &1 \\ 1&4\eem \;.
\ee

One can then verify that their inner product matrix, which will turn out to be the real root part of the generalized Cartan matrix of the algebra we are going to define in the next section, is given by (\ref{cartan2}). The fundamental Weyl chamber, defined by (\ref{funda_chamber_1}),
is therefore a regular square when projected on the Poincar\'e disk. See Figure \ref{Walls}.

By our construction, the Weyl group $W^{(2)}$ again plays a physical role as the group of wall-crossings or the group of the discrete attractor flow of the theory. One can verify that the four generators of the corresponding Weyl group $W^{(2)}$ (\ref{generator_weyl}) have no relations among themselves other than that $(s_i^{\scriptscriptstyle{(2)}})^2 = {\mathds {1}}$. This is an important consistency condition since this property ensures that there is a unique ``shortest-path ordering", the so-called weak Bruhat ordering, among the group elements, a property that an attractor flow always exhibits. See \cite{Cheng:2008fc} for details about the ordering and its precise relationship to the attractor flow.

The S-duality group of the theory is related to the Weyl group in a similar way as in the un-orbifolded case (\ref{S_Weyl_group}), namely
\be
\til\G_1(2)=W^{\scriptscriptstyle{(2)}}\rtimes \text{\small Sym}({\cal W}^{\scriptscriptstyle{(2)}})\;,
\ee
where the group $\text{\small Sym}({\cal W}^{\scriptscriptstyle{(2)}})$ is the group of symmetry of the fundamental square ${\cal W}^{\SS (2)}$ which is compatible with the lattice structure (\ref{quant}). In this case it is the four-element group generated by the order-two reflection element
\be
TR:  \a^{\scriptscriptstyle{(2)}}_i \leftrightarrow \a^{\scriptscriptstyle{(2)}}_{3-i\text{   mod   }4}
\ee
and the order-two next-next-neighbor rotation
\be
\g^{\scriptscriptstyle{(2)}}: \a^{\scriptscriptstyle{(2)}}_i \to
\a^{\scriptscriptstyle{(2)}}_{i-2 \text{   mod   }4}\;.
\ee

\subsubsection{$N=3$}

A very similar story holds for the $N=3$ case.
For the $\Z_3$ orbifold theory, the third row of the Stern-Brocot tree gives us six simple real roots (\ref{middle_tree}). Apart from $\a^{\scriptscriptstyle (3)}_1=\a_1$ and $\a^{\scriptscriptstyle (3)}_2=\a_2$, they are
\bea
\a_3^{(3)} &= \bem 4 &5 \\ 5&6\eem\;\;,\;\;\;
\a_4^{(3)} &=\bem 4 &7 \\ 7&12\eem \\ \notag
\a_5^{(3)} &= \bem 2 & 5\\ 5&12\eem \;\;,\;\;\;
\a_6^{(3)} &= \bem 0 &1 \\ 1&6\eem \;.
\eea
Their matrix of the inner products is given as (\ref{cartan3}). From this we can see that they define a fundamental Weyl chamber ${\cal W}^{\scriptscriptstyle (3)}$ which is a regular hexagon, and a Weyl group $W^{\scriptscriptstyle (3)}$ which is generated by six generators
$s_i^{\scriptscriptstyle (3)}$ with no relations other than $(s_i^{\scriptscriptstyle (3)})^2 = {\mathds 1}$. See Figure \ref{Walls}.
The S-duality group of the theory is
\be
\til\G_1(3)=W^{\scriptscriptstyle{(3)}}\rtimes \text{\small Sym}({\cal W}^{\scriptscriptstyle{(3)}})\;,
\ee
where the group $\text{\small Sym}({\cal W}^{\scriptscriptstyle{(3)}})$ is the group of symmetry of the fundamental hexagon, generated by again the order-two reflection element
\be
TR:  \a^{\scriptscriptstyle{(3)}}_i \leftrightarrow \a^{\scriptscriptstyle{(3)}}_{3-i\text{   mod   }6}
\ee
and the order-three next-next-neighbor rotation
\be
\g^{\scriptscriptstyle{(3)}}: \a^{\scriptscriptstyle{(3)}}_i \to
\a^{\scriptscriptstyle{(3)}}_{i-2 \text{   mod   }6}\;.
\ee
\section{Exact Partition Function for CHL Dyons }
\label{Exact}

In this section, we will start by introducing
the partition function for supersymmetric dyons in the $\Z_{\SS N}$ CHL models with $N=1,2,3$ as proposed in \cite{Dijkgraaf:1996it,Jatkar:2005bh}. We then see how they are related to an automorphic form of weight
\be\label{weight_Phit}
t=\frac{12}{N+1} -1
\ee
of a congruence subgroup $G_0(N)$ of the modular group $Sp(2,\Z)$.
In the second subsection we give explicit expressions for this automorphic form $\F_t(\O)$ both as an infinite product and as an infinite Fourier sum. These properties will enable us to construct a Borcherds-Kac-Moody algebra for each $N$ using these automorphic forms.

\subsection{The Dyon Partition Function}

In a $\mathbb{Z}_{\SS N}$ CHL model, the degeneracy of dyons  with given charges $(Q,P)$ and at a given point of the moduli space, whose coordinates are given by $\m$ and $\l$, is given by the following contour integral \cite{Dijkgraaf:1996it,Jatkar:2005bh}
\be\label{PF0}
D^{\SS{ (N)}}(Q, P)\lvert_{\m,\l} \,= (-1)^{Q\cdot P+1}\frac{1}{N}\,\oint_{\cal C} d\O \,\frac{e^{\p i (\Lambda_{Q,P}. \Omega)} }{(\F_{t}(\O))^2}\quad,\quad \O = \bem\r & \n \\ \n & \s\eem
\ee
The  integration contour is given by \cite{Cheng:2007ch}
\be\label{contour}
{\cal C}= {\cal C}(Q, P)\lvert_{\m,\l} = \{\im \O = \varepsilon^{-1} {\cal Z},\,0\leq \re\r, \tfrac{1}{N}\re\s,\re \n < 1\},
\ee
where $\varepsilon \ll 1$ is any small positive number playing the role of a regulator. For a given set of charges, the contour depends on the moduli $\mu, \lambda$ through the definition of the central charge vector (\ref{def_Z_vec}).
A remarkable property of the spectrum of ${\cal N} =4$ dyons is that the  entire moduli dependence of the degeneracy is captured completely by the moduli dependence of the choice of the contour.

To understand the properties of the above partition function,
let us recall a few facts about the congruence subgroups of $Sp(2, \mathbb{Z})$ and their Siegel modular forms. By definition, an $Sp(2, \mathbb{Z})$ element can be represented by a $(4\times 4)$ matrix which leaves invariant the symplectic form
\be
 J= \bem 0 & -{\mathds 1 }\\  {\mathds 1 } &0\eem\;.
\ee
When expressed in terms of ($2\times 2$) blocks, they are
\be
\bem A & B \\ C & D\eem \;\;\;,\text{   with} \;\;\;\;AB^T =BA^T,\;CD^T =DC^T,\;AD^T -BC^T =\mathds 1
\ee
and they act on the matrix of chemical potentials for the charge vector $\L_{Q,P}$, or the ``period matrix", as
\be
\O \to (A\O+B)(C\O+D)^{-1}\;.
\ee
The following subgroup of $Sp(2, \mathbb{Z})$, which we denote by $G_0(N)$, will be of special interest to us. In terms of matrices, these are elements satisfying the extra condition that they have the form
\be\label{automorphic_group}
U_0 \bem A & B \\ C &D\eem U_0^{-1}\;\;,\;\;\; C = \bem 0& 0\\0& 0\eem \text{  mod  }N\;,
\ee
where
\be
U_0 =\bem 1&0&0&0\\ 0
&0&0&1 \\ 0&0&1&0 \\ 0&-1&0&0\eem
\ee
can be thought of an $Sp(2,\Z)$ counterpart of $S= \big(\begin{smallmatrix} 0&1\\-1&0\end{smallmatrix}\big)$ in $SL(2,\Z)\sim Sp(1,\Z)$.
A special family of elements in $G_0(N)$ is given when one takes
\be
 \bem A & B \\ C &D\eem  = \bem a&  0&0 &b\\ 0&a&b& 0 \\ 0&c&d&0 \\ c& 0&0& d\eem\;,\;\; \g = \bem a & b\\ c& d\eem \in \G_0(N)\;,
\ee
then one can easily check that the corresponding $G_0(N)$ element acts on the period matrix as
\be
\O \to \g \,\O\, \g^T
\ee
and this gives a natural embedding of $\G_0(N) \subset SL(2,\Z)$ into  $G_0(N) \subset Sp(2,\Z)$.

For the partition function to converge, the matrix $\O$ has to lie in the so-called Siegel upper-half plane, given by the condition
$\Im \O \in V^+$. Furthermore, the object $\F_t(\O)$ turns out to be an automorphic form of weight $t$ under the subgroup $G_0(N)$ of $Sp(2,\Z)$, namely they transform as
\be\label{transformation_phi}
 \Phi_t [(A \Omega + B )(C\Omega + D ) ^{-1}] = \pm  \{\det{(C\Omega + D )}\}^{t}\,
    \Phi_t (\Omega)
\ee
when $\big(\begin{smallmatrix}A&B\\C&D\end{smallmatrix}\big)\in G_0(N)$. In particular, it is independently invariant under the following three transformations \be
\r \to \r+ 1\; ,\;\; \n\to \n+1\;,\;\; \s \to \s +N\;.
\ee
This invariance explains the real part of the contour of integration in (\ref{contour}).

This mathematical property of the automorphism of the partition function turns out to be closely related to the physics of crossing the walls of marginal stability in the moduli space.  The degeneracy formula as we wrote them in (\ref{PF0}), is moduli-dependent through the moduli dependence of the contour of integration. With such a contour prescription (\ref{contour}), changing the moduli causes a deformation of the contour.
As long as the contour does not encounter a pole of the partition function, the degeneracy does not change and hence is a smooth function in a region of moduli space. However, the partition function $1/(\F_t)^2$ is not entirely holomorphic and has poles in the Siegel upper-half plane. It turns out that the class of poles that one can hit when deforming the contour are in one-to-one correspondence with the two-centered solutions that can decay and are related to each other by $\til\G_0(N)$ transformations as we have seen in the last section.
Indeed, crossing a wall in the moduli space corresponds to crossing a pole in deforming a contour in accordance with the change of moduli  so the degeneracy jumps in a way consistent with the macroscopic analysis \cite{Sen:2007vb,Dabholkar:2007vk, Cheng:2007ch}. What is remarkable with these $\CN =4$ theories is that even though the degeneracy jumps from one region to another, it is possible to define a partition function globally over the moduli space such that its pole structure completely captures the intricate structure of walls of marginal stability.

As is suggested by our choice of notation, it turns out that in all cases of our interest, $N = 1, 2, 3$, the inverse partition function $(\Phi_t(\O))^2$ is a complete square of the automorphic form $\F_t(\O)$ of integral weight
$t$
of the congruence subgroup $G_0(N)$ of $Sp(2,\Z)$\footnote{It is not always true that the inverse partition function is a square of some automorphic form with integral weight. In particular, for $N=7$ the partition function cannot be written as the square of some infinite product with integral exponents.}. In particular, taking $A=(D^{-1})^T = \g \in\G_0(N)$ in (\ref{transformation_phi}), we see that it is invariant or anti-invariant (invariant up to a minus sign) under the transformation
$\O \to \g(\O), \,\g \in\G_0(N)$.
It is this form $\Phi_t$, with $t = 5, 3, 2$ for $N=1,2,3$, which will enable us to construct the Borcherds-Kac-Moody algebra underlying the dyon spectrum in $\S${\ref{Algebra}}.

\subsection{The Siegel Modular Forms}

In order for us to construct the generalized Kac-Moody algebra relevant for the dyon spectrum of the CHL model, and to be have an explicit knowledge about the set of imaginary simple roots and the multiplicities of all positive roots, it is crucial that we have the knowledge of the denominator of its characters both as an infinite product and as an infinite sum.  The automorphic form $\F_t(\O)$ discussed above is known to have an infinite product presentation, which can be derived using the type IIB realization of the dyons as D1-D5-momenta system in the KK-monopole background \cite{David:2006ji,David:2006yn}. The exponents in the product formula (\ref{PF_big}) are the Fourier coefficients of certain weak Jacobi forms $\chi^{(n,m)}(\t,z)$ of zero weight and index $1$ of the congruence subgroup $\G_0(N)$ of $SL(2,\Z)$
\be\label{root_mul_jacobi}
\chi^{\scriptscriptstyle{(n,m)}}(\t,z) = \sum_{\substack{k,\ell\in \Z\\ k>0}}\,c^{\scriptscriptstyle{(n,m)}}(\tfrac{4k}{N}-\ell^2)\,q^{k/{N}}y^\ell\;\;,\;\;q=e^{2\p i \t},\,y=e^{2\p i z}\;.
\ee
 These weak Jacobi forms are periodic with period $N$ in both $n$ and $m$ (\ref{modN}).
Physically, these weak Jacobi forms $\chi^{(n,m)}(\t,z)$ are related to the orbifold-invariant part of the elliptic genus of the two-dimensional CFT with target space $K3/\Z_{\scriptscriptstyle N}$, restricted to the $n'$-th twisted sector, where $0\leq n' < N$ and $n'=n$ mod $N$.
The explicit expression of them and in particular the proof of the integrality of their Fourier coefficients can be found in Appendix \ref{Root Multiplicities from Weak Jacobi Forms}. This relation between the product representation of an automorphic form and a modular or weak Jacobi form is sometimes known as the ``multiplicative lift" or the ``Borcherds lift"\cite{Borcherds:1995si,Gritsenko:1997au,Gritsenko:1996-2}.

Moreover, the automorphic form $\F_t(\O)$ has also a simple Fourier sum, with the Fourier coefficients given by the Fourier coefficients of the Jacobi form
\bea\label{sum_seed}
\f_{t,1/2}(\t,z) = i \, \eta^{\frac{12}{N+1}-3}(\t)\, \eta^{\frac{12}{N+1}}({\t}/{N})\,\theta_{1,1}(\t,z) =  \sum_{\substack{k>0\\ k,\ell \in \Z}} C(k,\ell)\, q^{\frac{k}{2N}} y^{\frac{\ell}{2}}
\eea
of the same weight $t$ and index $1/2$ of a subgroup of $SL(2,\Z)$ which is related to an S-duality group $\G_1(N)$ by a conjugation by the S-transformation (\ref{sum_seed_transf}).  This relation between the Fourier coefficients of the automorphic forms and those of a modular form is sometimes called the ``additive lift" or the ``arithmetic lift" \cite{maass,Gritsenko:1995ek}.

More explicitly, we have  the following two expressions for the automorphic form $\F_t(\O)$ as an infinite sum and as an infinite product
\bea\notag
\F_t(\O)&=&e^{2 \p i (\frac{1}{2N}\s+\frac{1}{2}\r+\frac{1}{2}\n)}\, \prod_{\substack{n,m,\ell\in\Z\\ (n,m,\ell)>0}} \big(1-
e^{2 \p i (\frac{n}{N}\s+m\r+\ell\n)}\big)^{c^{\scriptscriptstyle{(n,m)}}(\frac{4nm}{N}-\ell^2)}\\\label{PF_big}
&=&\sum_{\substack{\til n, \til m, \til\ell\in2\Z+1\\ \til n, \til m>0}}e^{2 \p i (\frac{\til n}{2N}\s+\frac{\til m}{2}\r+\frac{\til\ell}{2}\n)}\sum_{\substack{\d \lvert(\til n, \til m, \til\ell) \\ \d> 0}} \d^{t-1}\,  d(\d)\;C\big(\frac{\til n \til m}{\d^2},\frac{\til\ell}{\d}\big)\;,
\eea
where the Dirichlet characters in the sum formula are $d(\d)=1$ for $N=1$,
\be
d(\d)=\begin{cases} 0 &,\,\d = 0 {\text{   mod    }} 2\\ 1&,\,\d = 1 {\text{   mod    }} 2\end{cases}
\ee
for $N=2$ and
\be
d(\d)=\begin{cases} 0 &,\,\d = 0 {\text{   mod    }} 3\\ 1&,\,\d = 1 {\text{   mod    }} 3\\ -1&,\,\d = -1 {\text{   mod    }} 3\end{cases}
\ee
for $N=3$.
In section \ref{subsection_denominator} we will see how the above formula encodes all the information about the simple roots and the resulting root multiplicities of the algebra for CHL dyons.

\section{The Algebra for CHL Dyons }\label{Algebra}

The objective of this section is to first construct the generalized
Kac-Moody algebras from the partition functions of the CHL dyons and subsequently elucidate the role of this algebra in the supersymmetric dyon spectrum of the CHL models. To achieve this we first analyze the Weyl vector of the algebra and explain how the physical
``niceness" condition on the walls of marginal stability is translated into the mathematical ``niceness" condition of the existence of a time-like Weyl vector of \cite{Gritsenko:1996-1,Gritsenko:1996-2}. We then see, for models satisfying these ``niceness" conditions, the sum and product representations of the automorphic form $\F_t(\O)$ in the last section gives us the denominator identity of the algebra and therefore provides us with complete knowledge of both the simple roots as well as all the root multiplicities. Finally we see how the dyon spectrum furnishes a representation of the algebra and in what sense the algebra can been seen as an extra symmetry in the supersymmetric sector of the CHL theories. %The readers can consult Appendix \ref{review_BKM} for a summary of the relevant background knowledge about the generalized (or Borcherds-)Kac-Moody algebra.

\subsection{The Weyl Vector}
\label{The Weyl Vector}
One of the important objects in the theory of finite and affine Lie algebra is the Weyl vector. The Weyl vector of a Borcherds-Kac-Moody algebra is defined analogously as the vector in the root space which has inner product $-1$ with all simple real roots of the algebra
\be\label{def_weyl_vector}
(\varrho,\a_i) = -\frac{1}{2} (\a_i,\a_i) \quad\text{for all simple real roots} \quad \a_i\,.
\ee
In the present physical context, the Weyl vector is directly related to the level-matching condition of the orbifolds. Mathematically it is crucial for the automorphic properties of the partition function which will play an important role in the definition of the algebra. In this subsection we will see how this Weyl vector carries information about the topology of the system of walls of marginal stability of the theory. In particular, only if this vector is time-like, the theory will belong the finite class of models discussed in $\S$\ref{Three Arguments for Two Classes of Theories}.

To begin with, from the above definition of the Weyl vector, we expect this vector to be invariant under the action of elements of the symmetry group $\text{\small Sym}({\cal W}^{(N)})$
of the fundamental Weyl chamber, and in particular under the action of  the $\gamma^{{\scriptscriptstyle(N)}}$ given in
(\ref {sym_weyl_chamber}). This consistency condition actually fixes for us the Weyl vector up to a normalization factor and we are left with the following unique choice
\be\label{weyl_vector_1}
\varrho^{{\scriptscriptstyle(N)}}= \bem 1/N &1/2 \\ 1/2 & 1\eem\;.
\ee
Notice that this vector is space-like, light-like, or time-like exactly when the symmetry generator $\g^{{\scriptscriptstyle(N)}}$ is hyperbolic, parabolic or elliptic respectively. To understand this, recall that there is a map between a complex number $z$ in the interior of the upper-half plane to a ray in the future light-cone Minkowski space $\R^{2,1}$ by
\be
v(z) \sim\frac{1}{\im z} \bem |z|^2 & \Re z \\ \Re z& 1\eem\;,\;\;\Im z>0
\ee
such that the vector transforms as
\be\label{vector_transformation}
v(\frac{az+b}{cz+d}) = \g\,v(z)\,\g^T\;\;\;,\;\;\;\g = \bem a& b \\ c & d\eem\;.
\ee
This is the familiar map (\ref{ST_moduli}) which is often used to write down a manifestly S-duality action for supergravity, see for example \cite{Bergshoeff:1995sq}.
Furthermore, we have seen in $\S$\ref{Three Arguments for Two Classes of Theories} how this map can be extended to a map between the rational numbers $z=B/A, g.c.d.(A,B)=1$ on the boundary of the upper-half plane ${\cal H}$ and the boundary of the future light-cone, given by
\be
v({\frac{B}{A}}) \sim \bem B^2 &AB \\ AB& A^2\eem\;
\ee
such that the same transformation rule (\ref{vector_transformation}) applies. Now the question is, what about the rest of the real line, which is the boundary of the upper-half plane? Consider complex numbers of the form
\be
z= q_1 + q_2 \sqrt{D}\;,\;\;q_1,q_2 \in \Q
\ee
where $D$ is some square-free integer, notice that the transformed number $\frac{az+b}{cz+d}$ is again of this above form. This is why these numbers are sometimes said to be in the ``quadratic number field" $\Q(\sqrt{D})$.
Now we can write down the following map of these numbers to the rays in $\R^{2,1}$
\be
v(z) \sim \frac{1}{|q_2\sqrt{D}|} \bem q_1^2 - q_2^2 D & q_1 \\ q_1 & 1
\eem
\ee
such that the same transformation rule (\ref{vector_transformation}) again applies and the vector is time- or space-like depending on whether the integer $D$ is negative or positive.

Finally, not surprisingly the Weyl vectors are simply the image of the fixed point of the M\"obius transformation (\ref{mobius}) given by the symmetry transformation $\g^{\scriptscriptstyle{(N)}}$ under the above map.
Recall that we have shown that the number of walls are infinite for $N\geq 4$ because the symmetry generator $\g^{\scriptscriptstyle{(N)}}$ has real fixed points, we see that the infinitely many walls and the absence of a time-like Weyl vector is really one and the same thing. Therefore, the physical requirement that each domain in the moduli space should be bounded by a finite number of walls leads us to consider the root lattice admitting a time-like Weyl vector as discussed in  \cite{Gritsenko:1996-1,Gritsenko:1996-2}.

\subsection{Partition Function as a Denominator Formula}
\label{subsection_denominator}

In this section we would like to answer the following question: can the automorphic form $\F_t(\O)$ appearing in the dyon degeneracy formula for the CHL models (\ref{PF0}) be used to define a ``automorphic form corrected" Borcherds-Kac-Moody algebra? As announced in $\S$\ref{Summary}, the answer is positive for $\Z_{\SS N}$ models with $N<4$ and negative for the $N>4$ model. The obstruction for the $N>4$ case, whose partition function has also been proposed in \cite{Jatkar:2005bh}, basically lies in the fact that the would-be Weyl vector is space-like which brings in other inconsistencies with it.  This is in turn related to the property of the physical theory that there is no domain in the moduli space bounded by finitely many walls of marginal stability. We will focus in this section on the $N<4$ cases and see how the automorphic forms $\F_t(\O)$ specify all the data of the algebra for us. Our analysis will be analogous to the work of Gritsenko and Nikulin\cite{Gritsenko:1997au} in which $\F_5(\O)$ was used to construct a  Borcherds-Kac-Moody superalgebra.

A Borcherds-Kac-Moody superalgebra is most conveniently defined using its Chevalley basis, subject to a set of (anti-)commutation relations specified by the Cartan matrix. Therefore, we can conveniently decompose it into $\mathfrak{g}= \sum_{\a \in \D^+} \mathfrak{g}_{-\a} \oplus \mathfrak{h} \oplus \sum_{\a \in \D^+} \mathfrak{g}_{\a} $ where $\D^+$ denotes the set of positive roots. We will not write down their definitions nor discuss their properties here. The reader might find them in, for example, \cite{Ray,Gritsenko:1997au,Cheng:2008fc}.

Here we are in particular interested in the (super-) denominator identity of such a superalgebra, which reads
\be\label{denominator}
e(-\varrho) \prod_{\a \in  \D_+} (1-e(-\alpha)\,)^{\text{mult}\a}
=\sum_{w\in W} \, \text{\small det}(w) \, w\bigl( e(-\varrho) \,\Sigma\bigr)\;,
\ee
where $ \text{\small det}(w)=1$ (-1) if the group element $w$ can be written as an even (odd) number of reflections, and {\small mult}$\a$ denotes the graded (fermionic root weighted with $-1$) multiplicity of the positive root $\a$. When the algebra has imaginary simple roots (and therefore is ``generalized" or ``Borcherds-"), the sum involves the following correction term given by
\be
\Sigma= \sum_s \epsilon(s)\, e(-s)\,,
\ee
where $s$ denotes a sum of {\it imaginary} simple roots, $\ \epsilon(s) = (-1)^{n_{\bar{0}}}$ if $s$ is a sum of $n_{\bar{0}}$ number of pairwise perpendicular even (bosonic) imaginary simple roots and $n_{\bar{1}}$ number of odd (fermionic) imaginary simple roots, which are all distinct unless it is lightlike, and $\epsilon(s)=0$ otherwise. In the above formula, we have used the abstract exponentials $e(\a)$ satisfying the relation $e(\a)e(\a')=e(\a+\a')$\footnote{These abstract exponentials can be thought of as linear functions on the complexified root space. In our case it maps the space of symmetric $2\times 2$ complex matrices to $\C$ by $e(\a): v \mapsto e^{(\a, v)}$, where the exponential on the right-hand side is the ordinary one. Choosing $v = \pi i \O$ gives a `specialization'  of the denominator formula (\ref{denominator}) of the form (\ref{PF_denominator}) in our context.}.

In the present case, the vector space where the roots live is of a Lorentzian signature with only one time-like direction. Using the fact that no two vectors in the future light-cone are perpendicular to each other, we can rewrite the above equation as
\be\label{denominator2}
e(-\varrho) \prod_{\a \in  \D_+} (1-e(-\alpha)\,)^{\text{mult}\a}
=\sum_{w\in W} \, \text{\small det}(w) \,\bigg( e\bigl(w(\varrho)\bigr) - \sum_{\a\in{\cal W}} M(\a)\,e\bigl(w(\varrho+\a)\bigr) \bigg)\;,
\ee
where ${\cal W}$ again denotes the fundamental Weyl chamber.
The Fourier coefficients $M(\a)$ contain all the information of the imaginary simple roots of the algebra as summarized in Table \ref{imaginary_r_table}. The graded degeneracy of a simple root $\a$ is the number of times it is repeated in the set of simple roots  counted with plus/minus sign for $\a$ in the set of bosonic/fermionic simple roots. It is just given by the Fourier coefficients $M(\a)$ for time-like simple roots. For light-like ones it is given by
 the integers ${\til M}(\a)$, related to $M(\a)$ by
\be\label{lightlike_def}
1- \sum_{n \in \N} M(n\a) q^{n} = \prod_{k \in \N}(1-q^{k})^{\til M(k \a)}\;.
\ee

\begin{table}
\caption{\small{The List of Imaginary Simple Roots}}
\vspace{5pt}
\centering
\begin{tabular}{ccc}
\hline\vspace{1pt}
category  \Top\Bottom& condition& graded degeneracy\\
\hline
bosonic time-like  \Top\Bottom& $ (\a,\a) < 0,\;M(\a) > 0$ & $ M(\a) $ \\
\vspace{1pt}
bosonic light-like & $ (\a,\a) = 0,\;M(\a) > 0$ & ${\til M}(\a)$ \\
\vspace{1pt}
fermionic time-like & $ (\a,\a) < 0,\;M(\a) < 0$ & $M(\a)$ \\
\vspace{1pt}
fermionic light-like  \Top\Bottom& $ (\a,\a) = 0,\;M(\a) < 0$ & ${\til M}(\a)$ \\
\hline
\label{imaginary_r_table}
\end{tabular}
\end{table}

We would like to argue that the sum and product representations of the automorphic form $\F_t(\O)$ (\ref{PF_big}) give the right-hand side and the left-hand side of the above denominator identity (\ref{denominator2}) respectively, and therefore defines for us the Borcherds-Kac-Moody algebras we are looking for. If true, this means that the Fourier coefficients of the Jacobi forms $\chi^{\scriptscriptstyle{(n,m)}}(\t,z)$ and $\f_{t,1/2}(\t,z)$ encode the information about the root multiplicities and the list of imaginary simple roots respectively.
For this interpretation to be correct, the following consistency conditions must be satisfied (\ref{PF_big}) which are easy to verify.
%\hfill\eject
\begin{myitemize}
\item On  the product side, we show that
\begin{myenumerate}
\item all  exponents $c^{\SS (n,m)}(\frac{4nm}{N}-\ell^2)$ are integers and therefore might be interpreted as the graded root multiplicities,
\item the product has the form of the product side of the denominator (\ref{denominator2}) with Weyl vector $\varrho = \varrho^{\SS (N)}$ as derived in (\ref{weyl_vector_1}) from symmetry consideration,
\item all positive real roots are of the form
(\ref{posi_root_N}) and therefore are indeed in one-to-one correspondence with the walls of marginal stability of the theory.  Therefore the corresponding real simple roots are indeed those derived in $\S$\ref{The Finite Cases} by analyzing the physical walls of marginal stability of the theory.
\end{myenumerate}
\item On the sum side, we show that
\begin{myenumerate}
\item all Fourier coefficients are integers and therefore might be interpreted as the degeneracies of the imaginary simple roots,
\item the automorphic form  has the correct transformation property $\F_t(w(\O))=\text{\small det}(w)\,\F_t(\O)$ for all elements $w$ of the Weyl group,
 \item the sum involves only vectors in the future light-cone $w(\varrho+\a) \in V^+$, with $\a$ being a vector in the fundamental Weyl chamber and the Weyl vector $\varrho$ again given by same vector $\varrho^{\SS (N)}$  (\ref{weyl_vector_1}) as in the product expression.
\end{myenumerate}
\end{myitemize}
The details of the proof of these statements can be found in Appendix \ref{Root Multiplicities from Weak Jacobi Forms}.

In other words, we see that the square root of the inverse partition function for the $N=1,2,3$ CHL model (\ref{PF_big}) can be expressed in terms of the data of a Borcherds-Kac-Moody algebra as follows
\bea\notag
\F_t(\O) 	&=& \sum_{w\in W} \text{\small det}(w)\, \left( e^{-\p i (w(\varrho),\O)}-\sum_{\a\in\D_s^{im}} M(\a) e^{-\p i(w(\varrho+\a),\O)}\right)\\\label{PF_denominator}
&=& e^{-\p i (\varrho,\O)}\prod_{\a\in \D_+} \bigg(1- e^{-\p i(\a,\O)}\bigg)^{\text{mult}\,\a}\;.
\eea
For the readability we have suppressed the label ``$N$" denoting the different models. For example, $\varrho$ stands for the Weyl vector $\varrho^{\scriptscriptstyle(N)}$ (\ref{weyl_vector_1}) of the algebra for a given value of $N$ and $W$ stands for its Weyl group $W^{\scriptscriptstyle{(N)}}$. Similarly $\D_{s}^{im}$ and $\D_+$ stands for the set of imaginary simple roots and the positive roots of the $N$-th algebra respectively.

Explicitly, writing the vector $\a$ in the root lattice in its components as follows
\be\label{root_lattice}
\a=\bem2n/N & \ell \\ \ell & 2m\eem\;,
\ee
then the multiplicities of the positive roots are given by the Fourier coefficients of the weak Jacobi form $\chi^{\scriptscriptstyle(n,m)}(\t,z)$
\be
\text{\small mult }\a =  c^{\scriptscriptstyle(n,m)}(|\a|^2) =
c^{\scriptscriptstyle(n,m)}(\tfrac{4nm}{N}-\ell^2) \\
\ee
and the degeneracies of imaginary simple roots are given by the Fourier coefficients of the  Jacobi form  $\f_{t,1/2}(\t,z)$:
\bea\label{more_multi}
M(\a)=- \sum_{\substack{\d\lvert(2n+1,2m+1,2\ell+1)\\\d>0}}\d^{t-1}\,d(\d)\,C(\tfrac{(2n+1)(2m+1)}{\d^2},\tfrac{2\ell+1}{\d})
\eea
The first few values of $\text{\small mult }\a $ and $M(\a)$ can be found in Appendix \ref{Root Multiplicities from Weak Jacobi Forms}.

\subsection{Dyon Spectrum as a Representation of the Algebra}

After having constructed a Borcherds-Kac-Moody algebra from the partition function of the CHL models and verified that the Weyl group structure is the group structure of crossing the walls of marginal stability in the theory, we would like to ask what is the role of the rest of the algebra in the physical theory.  It turns out that the algebra we constructed for $N=2,3$ orbifold theory plays a basically identical role in the CHL models as the algebra constructed by Gritsenko and Nikulin \cite{Gritsenko:1997au} in the un-orbifolded theory, as recently discussed in \cite{Cheng:2008fc}. We will thus be rather brief in this part of the discussion and refer the readers to  \cite{Cheng:2008fc} for proofs and details of various statements in this subsection.

First, recall that a Verma module of a Borcherds-Kac-Moody superalgebra is an infinite-dimensional representation with a highest weight $\L$. Its super-character is given by
\be\label{verma_module}
\text{sch}\,{\mathfrak M}(\L) = \frac{e(-\varrho+ \L )}{e(-\varrho) \prod_{\a\in \D^+} \big(1-e(-\a)\big)^{\text{mult}\,\a}}\;.
\ee
The integrand in the degeneracy formula (\ref{PF0}) can thus be identified with the square of the character of the Verma module with a charge- and moduli-dependent highest weight
\be\label{highest_weight}
\L_w =\varrho + \frac{1}{2}\,w^{-1} (\L_{Q,P})  \;,
\ee
where $w\in W$ is the Weyl group element given by the moduli such that ${\cal Z} \in w({\cal W})$. In particular, the highest weight is simply $\varrho + \frac{1}{2}\,\L_{Q,P}$ when the moduli are fixed at their attractor values. Only in this case is the Verma module a ``dominant weight module", meaning that the highest weight of the module is inside the fundamental Weyl chamber.
Then the effect of the contour integration in computing the dyon degeneracy (\ref{PF0}) is simply to compute the (graded) dimension of the zero-weight subspace of the Verma module. In other words, the degeneracy of dyons at a given moduli is simply the number of different ways the vector $2\L_w$ can be written as a sum of two copies of positive roots, or a ``second-quantized" degeneracy of the weight $2\L_w$.

Second, an attractor flow in a black hole background
towards the attractor point of the moduli is described by a sequence of Weyl group elements following the so-called ``weak Bruhat ordering" of the group. This means, starting from a given point in the moduli space ${\cal Z} \in w({\cal W})$, there is a natural RG-flow-like sequence of Weyl group elements $w=w_n \to w_{n-1} \to \dotsi \to w_0 = {\mathds 1}$ and the moduli vector  ${\cal Z}$  follows the path $w_n({\cal W}) \to w_{n-1} ({\cal W}) \to \dotsi \to w_1({\cal W})  \to {\cal W}$
along the attractor flow. The corresponding Verma modules form a sequence of modules which decreases in size and terminates only when the highest weight is dominant
\be
{\mathfrak M}(\L_{w_n}) \supset {\mathfrak M}(\L_{w_{n-1}}) \supset \dotsi \supset {\mathfrak M}(\L_{w_0=\mathds 1})\;.
\ee
Combined with the relationship between the Verma modules and the dyon degeneracies discussed in the previous paragraph, this sequence gives a prescription of how to compute the difference in dyon degeneracies when a wall of marginal stability is crossed. This prescription gives a microscopic derivation of the so-called wall-crossing formula for the present $\CN=4$ theories.

Finally, we would like to comment on the role of the Borcherds-Kac-Moody algebra we constructed as a symmetry of the supersymmetric dyon spectrum. As implied in our discussion about the Verma modules, the dyon spectrum is generated by a set of  freely-acting bosonic and fermionic oscillators, with each positive root of the algebra corresponding to two copies of such oscillators and the Weyl vector $\varrho$ stipulating the vacuum of the system. More explicitly, with a given choice of simple roots we can rewrite (\ref{PF0}) as
\be\label{PF_new}
\frac{1}{e(-2\varrho) \prod_{\a \in \D^+} \big(1 - e(-\a)\big)^{2\,\text{mult}\a} } = \sum_{w\in W,\til\L \in {\cal W}}D(w(\til\L))\,e(w(\til\L))\;,
\ee
and the coefficient $D(w(\til\L))$  denotes the degeneracy
\be
D(w(\til\L)) =D(Q,P)\lvert_{\m,\l} \,\,,\quad \til\L =\L_{Q,P}, \quad{\cal Z} \in w^{-1}({\cal W})\;.
\ee
Notice that the degeneracies of any other charges with $\L_{Q,P} \not\in {\cal W}$ can be obtained by a simultaneous S-duality transformation (\ref{extended_S-dual}) of the charges and the moduli of the formula above.
What the above formula implies is, by acting on a dyon microstate by an element of ${\mathfrak g}_\a$ one obtains another dyon state as long as $\a$ is a bosonic positive root. The same goes for fermionic positive roots except for the fact that one has to be more careful with the exclusion principle as usual. In this sense, the Borcherds-Kac-Moody algebra we constructed plays the role as a spectrum-generating symmetry of the BPS spectrum. While the physical relevance of the Weyl group symmetry has been elucidated in $\S$\ref{Discrete}, a physical understanding of this larger symmetry is yet to be developed.

\section{Discussion} \label{Discussion}

We have identified a Borcherds-Kac-Moody superalgebra underlying the spectrum of dyon in $T^6/\mathbb{Z}_N$ CHL orbifolds for $N=1, 2, 3$.
Analogous to the $N=1$ case of toroidal compactification discussed in \cite{Cheng:2008fc}, we conclude that the algebra we constructed in the present paper plays the following role in the physical theory. First, the Weyl group gives the underlying group structure of crossing the walls of marginal stability. Second, the spectrum is generated by a set of bosonic and fermionic freely-acting oscillators with charge- and moduli-dependent oscillation levels. In particular, this algebraic microscopic model provides an alternative derivation of the wall-crossing formula.

Two important questions remain unanswered. The first question is: what is the microscopic model for the $\Z_{\SS N}$ CHL models with $N\geq 4$? As we have seen, on the macroscopic side the walls of marginal stability of these theories do not satisfy the natural ``niceness" condition, namely they do not render a partition of the moduli space into domains bounded by finitely many walls. Related to this, we expect the relevant lattice to have a light- or space-like lattice Weyl vector \cite{Gritsenko:1996-1,Gritsenko:1996-2}. On the microscopic side, we conclude that the partition functions proposed in \cite{Jatkar:2005bh} for these models cannot be related to the denominator formula of some Borcherds-Kac-Moody algebra in a similar way as in the $N<4$ models. This fact seems to be related to the appearance of some very light states (more precisely, the so-called ``polar states" which are not heavy enough to form black holes) in the spectrum. By way of an answer, one can entertain several possibilities. Either the microscopic models exist but are much more involved than in the case of the finite models, or  these theories are non-perturbatively inconsistent in some yet unknown way, or the BKM symmetry is only a spectrum-generating symmetry which may or may not exist in a given model.

The second unanswered question is: how should we understand physically the appearance of these algebras underlying the dyon spectrum? What we have achieved, apart from constructing these new algebras, is that we have now understood the significance of all real roots of the algebra as corresponding to the multi-centered solutions which cause the jump in the spectrum (see also \cite{Cheng:2008fc}). Furthermore, we have given a prescription of how the ``highest weight" of the relevant Verma module (\ref{highest_weight}), or the ``oscillation level" in  the partition function (\ref{PF_new}), is given by the total charges and the moduli. This gives us a microscopic model of the supersymmetric dyons in the $\CN=4$ CHL models. We have also shown that the models are very likely to be correct, since they reproduce the asymptotic growth as expected from the semi-classical Bekenstein-Hawking entropy and all its known corrections, and, on a much finer level, they reproduce the wall-crossing formula as predicted from the supergravity analysis (see also \cite{LopesCardoso:2004xf,Dijkgraaf:1996it,Cheng:2008fc}). But we do not yet understood the physical origin of this Borcherds-Kac-Moody symmetry. Understanding this symmetry will lead to a far more  complete understanding of the non-perturbative supersymmetric states in these theories than has been possible in  other four-dimensional gravitational theories less supersymmetry.

Finally, having identified an algebra, it is natural to consider the symmetry obtained by the `exponentiation' of this algebra. One would like to know whether this symmetry is a global symmetry or a gauge symmetry, whether it is unbroken or broken, and whether it is a genuine symmetry or a spectrum-generating symmetry. Clearly, the symmetry includes the Weyl group which is essentially the physical duality group. Now, at a generic point in the moduli space,  duality symmetry is a  spontaneously broken discrete gauge symmetry to the extent that one identifies theories at different points related by duality.  Therefore, at least a part of this symmetry is a genuine, spontaneously broken, gauge symmetry. It would be interesting to better understand the nature  of the full symmetry and in particular whether it corresponds to a large spontaneously broken gauge symmetry.

\section*{Acknowledgments}

It is a pleasure to thank Frederik Denef, Valery Gritsenko, Axel Kleinschmidt, Andy Neitzke, Daniel Persson, Ashoke Sen, Erik Verlinde, Xi Yin for useful discussions. We would like to thank the ICTS and the organizers of the `Monsoon Workshop in String Theory' at the TIFR, and IPMU of the Tokyo University for hospitality where part of this research was completed. The work of
M.C. is supported by the Netherlands Organization for Scientific Research (NWO); and of A.~D. is supported in part by the Excellence Chair of the Agence Nationale de la Recherche (ANR).

\appendix

\section{Root Multiplicities from Weak Jacobi Forms}
\label{Root Multiplicities from Weak Jacobi Forms}
\subsection{The Product Representation and the Positive Roots}
\label{The Product Formula and the Positive Roots}

In the type IIB duality frame, the quarter-BPS dyons can be realized as bound states of D1-D5-P and the KK monopole together with momenta along the KK monopole circle. By quantizing the relevant degrees of freedom, the product expression (\ref{PF_big}) of the partition function $(\F_t(\O))^2$ has been derived in \cite{David:2006yn}. As mentioned in (\ref{root_mul_jacobi}), the exponents of the products, or the graded multiplicity of the positive roots (\ref{PF_denominator}), are given by the Fourier coefficients of the following weak Jacobi form
\be\label{weak_expression}
\chi^{\scriptscriptstyle{(n,m)}}(\t,z) = \frac{1}{2N} \sum_{s \in \Z \text{ mod }\scriptscriptstyle{N}}e^{2\p i s m } F^{\scriptscriptstyle{(n,s)}}(\t,z) =
\sum_{\substack{k,\ell\in \Z\\ k>0}}\,c^{\scriptscriptstyle{(n,m)}}(\tfrac{4k}{N}-\ell^2)\,q^{k/{N}}y^\ell\;,
\ee
where $F^{\scriptscriptstyle{(n,s)}}(\t,z)$ have the following interpretation in the D1-D5 CFT: it is the $n$-th twisted elliptic genus of the two-dimensional $(4,4)$ supersymmetric conformal field theory with target space $K3/\Z_{\scriptscriptstyle{N}}$, with the insertion of the orbifold generator to the $s$-th power. The sum $\frac{1}{N} \sum_{s \in \Z \text{ mod }\scriptscriptstyle{N}}e^{2\p i s m } F^{\scriptscriptstyle{(n,s)}}(\t,z) $ therefore projects out the states  that are not invariant under the orbifold action. From this we immediately see that the pair of integers $(n,m)$ in the above formula can be seen as only defined up to mod $N$:
\be\label{modN}
\chi^{\scriptscriptstyle{(n,m)}}(\t,z) =\chi^{\scriptscriptstyle{(n+N,m)}}(\t,z) =\chi^{\scriptscriptstyle{(n,m+N)}}(\t,z) \;.
\ee
The explicit expressions for $F^{\scriptscriptstyle{(n,s)}}(\t,z)$ can be found in \cite{David:2006yn}, using which we derive the following expression for the weak Jacobi forms $\chi^{\scriptscriptstyle{(n,m)}}(\t,z) $

\begin{itemize}
\item{$n,m = 0$  mod $N$}

\bea\notag
\chi^{\scriptscriptstyle{(n,m)}}(\t,z) &=& \frac{1}{N+1}\bigg(2\,\f_{0,1}(\t,z) + (N-1) \f_{-2,1}(\t,z)\,\f_2^{\scriptscriptstyle(N)}(\t)\bigg)\\
\label{expression_1}
&=&  \f_{-2,1}(\t,z) + \frac{24}{N+1} \bigg( \varphi(\t,z)
+ \f_{-2,1}(\t,z)\,\sum_{k=1}^\inf \s_1(k)\,(q^k-N q^{kN})
\bigg)
\eea
\item{$n$  or  $m = 0$  mod $N$}

\bea \notag
\chi^{\scriptscriptstyle{(n,m)}}(\t,z)
&=& \frac{1}{N+1}\bigg(\f_{0,1}(\t,z) - \f_{-2,1}(\t,z)\,\f_2^{\scriptscriptstyle(N)}(\t)\bigg)\\ \label{expression_2}
&=&  \frac{12}{N+1} \bigg( \varphi(\t,z)
-\frac{2}{N-1} \f_{-2,1}(\t,z)\,\sum_{k=1}^\inf \s_1(k)\,(q^k-N q^{kN})
\bigg)
\eea
\item{$n,m \neq 0$  mod $N$}
\bea \notag
\chi^{\scriptscriptstyle{(n,m)}}(\t,z) &=&
-\frac{1}{N(N+1)} \f_{-2,1}(\t,z)\,\sum_{k \in \Z \text{ mod } \scriptscriptstyle N}\,e^{-2\p i \frac{nmk}{\scriptscriptstyle N}}\f_2^{\scriptscriptstyle(N)}(\tfrac{\t+k}{N})\\ \label{expression_3}
&=&  -\frac{24}{N^2-1} \;\f_{-2,1}(\t,z) \sum_{k=nm \text{ mod }N} \s_1(k)\,q^{k/N}\;.
\eea
\end{itemize}
In the above formulas, $\f_2^{\scriptscriptstyle(N)}(\t)$ is a weight 2 modular form under the subgroup $\G_1(N)$ and \be\label{divisor}
\s_x(n) =\sum_{d\lvert n} d^x\ee denotes the divisor function. The precise expressions for the weak Jacobi forms and modular forms appearing in the above formula can be found in Appendix \ref{More on Weak Jacobi Forms}.

The first lines of the above expressions for $\chi^{\scriptscriptstyle{(n,m)}}(\t,z)$ makes it manifest that they are weak Jacobi forms of zero weight and index 1 with respect to the congruence subgroup $\G_1(N)$. In other words, from the modular invariance and the spectral flow invariance of the CFT partition functions $F^{\SS (n,m)}(\t,z)$, one can show that
 they satisfy the two defining conditions for weak Jacobi forms
\bea\notag
\chi^{\scriptscriptstyle{(n,m)}}(\frac{a\t+b}{c\t+d},\frac{z}{c\t+d}) &=& \e\, e^{2\p i\frac{c z^2}{c\t+d}} \,\chi^{(n,m)}(\t,z)\;\;\;,\;\e^N =1 \;\;,\quad \bem a&b\\c&d\eem  \in \G_1(N)\notag\\
\chi^{\scriptscriptstyle{(n,m)}}(\t,z+\ell\t+m)&=&e^{-2\p i(\ell^2\t+2\ell z)}\chi^{(n,m)}(\t,z)\;,\quad\ell,m\in \Z.
\eea
On the other hand, the second lines of the above expressions make it manifest that all the Fourier coefficients $c^{\scriptscriptstyle{(n,m)}}(\frac{4nm}{N}-\ell^2)$ are indeed all integers and are amenable to the interpretation as the graded multiplicity of the positive roots. See Appendix \ref{More on Weak Jacobi Forms} for a detailed proof of the integrality property.

Now we can straightforwardly read out the (graded) root multiplicity $c^{\scriptscriptstyle{(n,m)}}(|\a|^2)$ of positive roots.
The few lowest lying ones are listed in table \ref{table_root_mul}.

\begin{table}
\centering
\caption{Root multiplicity of root {$\a$} in the form (5.13)}
\vspace{4pt}
\begin{tabular}{cccccccc}
\hline
&\multicolumn{2}{c}{$N=1$}\Top\Bottom& \multicolumn{2}{c}{$N=2$}&\multicolumn{2}{c}{$N=3$}\\
\cline{2-7}
&{$|\a|^2$}\Top\Bottom&{\small mult}\,$\a$&{$|\a|^2$}&{\small mult\,$\a$}&{$|\a|^2$}&{\small mult$\,\a$}\\\hline
\multirow{6}*{$n, m=0$ mod $N$}\Top\Bottom&-1&1&-1&1&-1&1\\
&0&10&0&6&0&4\\
&3&-64&3&-32&3&-22\\
&4&108&4&52&4&36\\
&7&-513&7&-257&7&-171\\
&8&808&8&408&8&268\\
\hline
\multirow{6}*{$n$ or $m=0$ mod $N$}\Top\Bottom&&&0&4&0&3\\
&&&3&-32&3&-21\\
&&&4&56&4&36\\
&&&7&-256&7&-171\\
&&&8&400&8&270\\
&&&11&-1376&11&-918\\
\hline
\multirow{6}*{$n$, $m\neq0$ mod $N$}\Top\Bottom&&&1&-8&{1}/{3}&-3\\
&&&2&16&{4}/{3}&6\\
&&&3&-24&{5}/{3}&-9\\
&&&4&48&{8}/{3}&18\\
&&&5&-96&{3}&-12\\
\Top\Bottom&&&6&160&4&24\\
\hline
\end{tabular}
\label{table_root_mul}
\end{table}

In particular, we find that the positive real roots are exactly those vectors labeling the walls of marginal stability of the theory (\ref{posi_root_N}) with multiplicity one, and the light-like positive roots have the degeneracies
\be\label{lightlike_multi_1}
\text{\small mult}\,\a=\begin{cases} \frac{24}{N+1}-2 & \text{when } n,m = 0 \text{ mod }N \\
\frac{12}{N+1}& \text{when } n \text{ or }m = 0 \text{ mod }N \\
0& \text{otherwise }\;.
\end{cases}
\ee
Furthermore, when $N>1$, the lowest-lying fermionic roots have length $|\a|^2=|\varrho^{\SS (N)}|^2=\frac{4}{N}-1$ (time-like) and graded multiplicity $-\frac{24}{N^2-1}$.

\subsection{The Sum Representation and the Imaginary Simple Roots \label{The Sum Formula and the Imaginary Simple Roots}}

In this appendix we will analyze the Fourier sum expression of the automorphic form (\ref{PF_big}). First we will show that it can be written as the sum expression of the denominator formula of a Borcherds-Kac-Moody formula (\ref{denominator2}), and then discuss some properties of the imaginary simple roots following from such an expression.

As mentioned in (\ref{sum_seed}), the Fourier coefficients of the automorphic form $\F_t(\O)$ are given by the Fourier coefficients of the following weight $t$, index $1/2$ Jacobi form
\bea\notag
\f_{t,1/2}(\t,z)&=& i \eta^{\frac{12}{N+1}-3}(\t)
\eta^{\frac{12}{N+1}}({\t}/{\textstyle N})\,\theta_{1,1}(\t,z) \\ \notag
&=&q^{\frac{1}{2N}}y^{\frac{1}{2}} \prod_{n\geq 1} (1-q^n)^{\frac{12}{N+1}-2}
(1-q^{n/N})^{\frac{12}{N+1}} (1-q^n y )(1-q^{n-1} y^{-1} )\\\label{sum_seed_app}
&=& \sum_{\substack{k>0\\ k,\ell \in \Z}} C(k,\ell)\, q^{\frac{k}{2N}} y^{\frac{\ell}{2}}\;.
\eea
It is a Jacobi form with respect to the congruence subgroup $\G_0(N)$ conjugated by the $S$-transformation. Namely, we have
\be\label{sum_seed_transf}
\f_{t,1/2}\big(\,\frac{a\t+b}{c\t+d},\frac{z}{c\t+d}\,\big) = e^{\p i 	 \frac{z^2}{c\t+d}}\,\,[-i (c\t+d)]^t\,\f_{t,1/2}(\t,z)\;\;,\;
\ee
if
\be\;\bem a & b \\c & d \eem\in PSL(2,\Z)\;,\;\;b= 0 \text{ mod }N\;.
\ee

As mentioned in the text before formula (\ref{PF_denominator}), there are three conditions we should check for the sum formula of the automorphic form to have an interpretation as the sum formula in the denominator identity of certain generalized Kac-Moody algebra.

First of all, the integrality of the Fourier coefficients $C(k,\ell)$ is manifest in the above definition (\ref{sum_seed_app}) of the Jacobi form $\f_{t,1/2}(\t,z)$. Second, the property of the theta-function
\be
\theta_{1,1}(\t,-z)=-\theta_{1,1}(\t,z)
\ee
translates into the transformation property
\be
\F_t(s_1(\O)) = - \F_t(\O)
\ee
of the automorphic form $\F_t$, where $s_1$ denotes the reflection with respect to the simple real root $\a_1$ (\ref{simple_real_root_1}). Conjugated with the symmetry generator of the fundamental Weyl chamber, for example $(\g^{\SS (N)})^{-1} s_1 \g^{\SS (N)} =s_3^{\SS (N)}$, we get
\be
\F_t(s_i(\O)) = - \F_t(\O)\;.
\ee
That is to say, the automorphic form is anti-invariant under any Weyl reflection (\ref{generator_weyl}). The composition of more than one reflections then gives the transformation property of the automorphic form $\F_t(\O)$ under  the action of a general Weyl group element
\be\label{transformation_sum_formula}
\F_t(w(\O)) =  \text{\small det}(w) \F_t(\O) \;.
\ee
This is the property that enables us to write the automorphic form as a graded sum of the images of the same thing under the Weyl group action as in the denominator formula (\ref{denominator}).
Third, from (\ref{sum_seed_app})
it is easy to check that the Fourier coefficients have the property
\be
C(k,\ell) = 0 \;\;\text{     if     }\;\;\frac{4k}{N}-\ell^2 < \rvert \varrho^{\scriptscriptstyle(N)}\lvert^2 = \frac{1}{N} - \frac{1}{4}\;.
\ee
In particular, the coefficients of the term $e^{-\p i (\b,\O)}$ in the Fourier expansion of $\F_t(\O)$ is non-zero only when $\b$ is inside the future light-cone. Using (\ref{transformation_sum_formula}) we can then concentrate on the part of the sum with $\b \in {\cal W}$. Notice that $\b$ cannot be on the boundary of any Weyl chamber, basically because the Weyl vector is not in the root lattice, and therefore $(\varrho+\a,\a')=1$ mod $2$ for any root $\a$, $\a'$. Now write $\b=\varrho+\a$, following the definition of the Weyl vector (\ref{def_weyl_vector}) we conclude that $\a$ must also lie within the fundamental Weyl chamber. This is the third condition we mentioned in $\S$\ref{subsection_denominator} for the sum expression for the automorphic form $\F_t(\O)$ to be written as the sum expression for the denominator of a certain Borcherds-Kac-Moody algebra.

It is now a straightforward task to read out the degeneracies of the imaginary simple roots we are interested in.  For example,
from the factors of the Dedekind eta functions in (\ref{sum_seed}) we can see that the multiplicities (\ref{lightlike_def}) of the light-like simple roots are
\be
\til M(k\a) = \begin{cases}
\frac{24}{N+1}-3 & \text{     when     }\; n,m = 0 \text{ mod }N \\
\frac{12}{N+1} & \text{     when     }\; n\text{ or }m = 0 \text{ mod }N
\\ 0 & \text{     otherwise     }
\end{cases}\;\;\quad, \quad\text{   for all   } k\in \N\;.
\ee
Comparing it with the multiplicity of the light-like positive roots $\a=
\big(\begin{smallmatrix}2n/N & \ell \\ \ell & 2m\end{smallmatrix}\big)$ we found in the previous subsection (\ref{lightlike_multi_1}), we see that the two are indeed consistent, taking into account the fact that the same vector $\a$ can also be written as a combination involving two real roots when $n,m=0$ mod $N$. For example, the light-like simple root $\big(\begin{smallmatrix} 2 & 2 \\ 2 & 2 \end{smallmatrix}\big) $ has degeneracy $9$, together with the possibility of writing it as the sum $\a_2+\a_3$ of two real simple roots, we conclude that the multiplicity of the positive root $\big(\begin{smallmatrix} 2 & 2 \\ 2 & 2 \end{smallmatrix}\big) $ is $10= c^{\SS (0,0)}(0)$.

As the next example, let us consider the multiplicity of the lowest lying fermionic roots $2\varrho^{\scriptscriptstyle(N)}$ in the $N>1$ cases. For the $N=2$ case, from $C({1},1)=1$, $C({9},3)=-1$ and using (\ref{more_multi}) we conclude that the multiplicity of this simple root should be $-(-1+3^2)= -8$. For $N=3$, from $C(9,3)=3$ we conclude that the multiplicity should be $-3$. This is indeed consistent with the result of a root multiplicity $-\frac{24}{N^2-1}$ from the product representation.

It is worth noting that from the Fourier expansion of (\ref{sum_seed_app}) that the numerical values of the degeneracies of the imaginary simple roots grow very slowly. They are  still of order $\lesssim 10^2$ when $|\a|^2 \sim 10^2$, in contrast with the root multiplicities encoded in the weak Jacobi form (\ref{weak_expression}), which are of order $\sim 10^{10}$ when  $|\a|^2 \sim 10^2$. The denominator identity  still holds thanks to the impressive cancelation between the contribution from the fermionic and the bosonic roots, which translates in the partition function into the large cancelation between the contribution from the fermionic and bosonic states to the index.

\subsection{More on Weak Jacobi Forms}
\label{More on Weak Jacobi Forms}

In this subsection we collect various definitions and formulas of the modular forms we have used in this paper. In particular we will explain the objects involved in the construction of the weak Jacobi forms $\chi^{\scriptscriptstyle{(n,m)}}(\t,z)$ in Appendix \ref{The Product Formula and the Positive Roots}, and prove the integrality of their Fourier coefficients.
For details of modular forms and weak Jacobi forms with respect to congruence subgroups, one can consult, for example, \cite{apostol,EicZ}.

Let us begin by recalling the two Eisenstein series and their relationship to the modular discriminant $\D(\t)$
\bea\notag
E_4(\t) &=&1 -\frac{8}{B_4} \sum_{n\geq 1} \s_3(n) q^n =  1 +240 \sum_{n\geq 1} \s_3(n) q^n \\ \notag
E_6(\t) &=&1 -\frac{12}{B_6} \sum_{n\geq 1} \s_5(n) q^n =  1 -504 \sum_{n\geq 1} \s_5(n) q^n\\\label{eisenstein}
E_4^3(\t) - E_6^2(\t) &=&1728 \,\D(\t) = 1728\, \eta^{24}(\t) \;,
\eea
where $\s_x(n)$ are again the divisor functions  defined as in (\ref{divisor}).

Using these and the their Jacobi form counterparts $E_{4,1}(\t,z)$ and $E_{6,1}(\t,z)$ (see \cite{EicZ}), we can write down the following weight $0$, weight $-2$ and index 1 weak Jacobi forms with which we have constructed our weak Jacobi forms $\chi^{\scriptscriptstyle{(n,m)}}(\t,z)$
\ben
\f_{-2,1}(\t,z) &=& -\eta^{-6}(\t)\,\th^2_1(\t,z) =  \frac{E_6(\t)E_{4,1}(\t,z)-E_4(\t)E_{6,1}(\t,z)}{144\D(\t)} \\
&=& (y^{-1}-2+ y) + (-2 y^{-2}+8 y^{-1}-12+8 y-2 y^2) q \\ &&+ (y^{-3}-12y^{-2}+39y^{-1}-56+39 y-12 y^2+y^3) q^2+ {\cal O}(q^3)\\
\f_{0,1}(\t,z) &=& \frac{\th^2_2(\t,z)}{\th^2_2(\t,0)}+ \frac{\th^2_3(\t,z)}{\th^2_3(\t,0)}+ \frac{\th^2_4(\t,z)}{\th^2_4(\t,0)} =  \frac{E_4^2(\t)E_{4,1}(\t,z)-E_6(\t)E_{6,1}(\t,z)}{144\D(\t)}\\
&=&(y^{-1}+10+y)+(10y^{-2}-64y^{-1}+108-64 y+10 y^2) q\\
&&+(y^{-3}+108y^{-2}-513y^{-1}+808-513 y+108 y^2+y^3) q^2
+ {\cal O}(q^3)\;,\een
These two weak Jacobi forms $\f_{-2,1}(\t,z) $ and $\f_{0,1}(\t,z)$,  together with the two Eisenstein series $E_4(\t)$, $E_6(\t)$ introduced above, generate the ring of weak Jacobi forms of even weight and arbitrary integral indices.

Another element we need is the following weight two modular form under the congruence subgroup $\G_0(N)$ when $N$ is prime
\ben
\f_2^{\scriptscriptstyle(N)}(\t) &=&
q\pa_q \log\bigg(\frac{\D(N\t)}{\D(\t)}\bigg)^{\frac{1}{N-1}}\\
&=& 1 +\frac{24}{N-1}  \sum_{k=1}^\inf \s_1(k)  \,(q^k - N q^{kN})\;.
\een

Using the modular properties of these modular and Jacobi forms we see that $\chi^{\scriptscriptstyle{(n,m)}}(\t,z)$ given in Appendix \ref{The Product Formula and the Positive Roots} are indeed weak Jacobi forms of weight zero and index one. But it is not clear whether they will have integral Fourier coefficients since they involve non-integral combinations of the above forms.

To show that they nevertheless have integral coefficients, let us rewrite them in a slightly different form as in (\ref{expression_1})-(\ref{expression_3}), where we have
defined the following object which does not have nice modular properties
\ben
\varphi(\t,z) &=& \frac{1}{12} (\f_{0,1}(\t,z) -\f_{-2,1}(\t,z) ) \\
&=&1+(y^{-2}-6y^{-1}+10-6 y+y^2) q\\
&&+(10y^{-2}-46y^{-1}+72-46 y+10 y^2) q^2+ {\cal O}(q^3)\;.
\een

To show that it has integral coefficients, notice that (\ref{eisenstein})
\be
\frac{1}{12} \big(E_4(\t)\f_{0,1} - E_6(\t) \f_{-2,1}(\t,z)\big) = E_{4,1}(\t,z)\;.
\ee
Since all the coefficients of $E_4(\t)$ and $E_6(\t)$ are divisible by 12 except for the constant terms (\ref{eisenstein}), namely $E_4(\t),E_6(\t)=1$ mod $12$, we conclude that $\varphi(\t,z) $ and therefore $\chi^{\scriptscriptstyle{(n,m)}}(\t,z)$ always have integral Fourier coefficients. This is of course a necessary condition for the product representation to have an interpretation as the denominator formula of a certain BKM algebra, as mentioned in the main text in $\S$\ref{subsection_denominator}.

\section{Notations and Definitions}

For the convenience of the readers, we collect the definitions of various objects frequently used in the main text.
\begin{itemize}

\item{Vectors in $\R^{2,1}$}

\begin{description}
\item[$\L_{Q,P}\;\;$]The vector of T-duality invariants. See (\ref{matrix_charge_vector}).
\item[${\cal Z}\;\;$]The central charge like vector encoding all the moduli dependence. See (\ref{def_Z_vec}).
\item[$\O\;\;$]The vector of chemical potentials/integration variables. See (\ref{PF0}).
\end{description}

\item{Groups}

\begin{description}
\item[$PSL(2,\Z)$] The S-duality group of the un-orbifolded theory.
\item[$PGL(2,\Z)$] The extended S-duality group of the un-orbifolded theory. See (\ref{pgl2z}).
\item[$\G_1(N)$] The S-duality group of the $\Z_{\SS N}$ orbifold theory. A subgroup of $PSL(2,\Z)$. See (\ref{Sgamma}).
\item[$\til\G_1(N)$] The extended S-duality group of the $\Z_{\SS N}$ orbifold theory. A subgroup of $PGL(2,\Z)$. See (\ref{Sgamma}).
\item[$\til\G_0(N)$] The group relating different walls of marginal stability of the $\Z_{\SS N}$ orbifold theory. A subgroup of $PGL(2,\Z)$. See (\ref{tilg0}). For $N<4$ we have $\til\G_0(N)=\til\G_1(N)$.
\item[$G_0(N)$] The $Sp(2,\Z)$ subgroup of which the Siegel modular form $\F_t(\O)$ is an automorphic form. See (\ref{automorphic_group}).

\end{description}

\item{Objects of the Algebra}

\begin{description}
\item[$\varrho\;\;$]The Weyl vector. See (\ref{def_weyl_vector}).
\item[$\varrho^{\SS (N)}\;\;$]The Weyl vector of the algebra for the $\Z_{\SS N}$ theory. See (\ref{weyl_vector_1}).
\item[$\D_+\;\;$] The set of all positive roots.
\item[$\D_s^{im}\;\;$] The set of all imaginary simple roots.
\item[${\mathfrak M}(\L)$] Verma module with highest weight $\L$. See (\ref{verma_module}).
\item[$\a_i^{\SS (N)}\;\;$] The $i$-th simple real root of the algebra for the $\Z_{\SS N}$ theory. See $\S$\ref{The Finite Cases}.
\item[$s_i^{\SS (N)}\;\;$] The reflection with respect to the $i$-th simple real root of the algebra for the $\Z_{\SS N}$ theory. See (\ref{generator_weyl}).
\item[${ W}^{\SS (N)}\;\;$] The Weyl group of the algebra for the $\Z_{\SS N}$ theory, which is the reflection group generated by $s_i^{\SS (N)}$.
\item[${\cal W}^{\SS (N)}\;\;$] The fundamental Weyl chamber of the algebra for the $\Z_{\SS N}$ theory. See (\ref{funda_chamber_1}).
\item[$\g^{\SS (N)}$] The generator of a particular symmetry of the fundamental Weyl chamber of the algebra for the $\Z_{\SS N}$ theory. See (\ref{sym_weyl_chamber}).
\end{description}
\end{itemize}
\bibliographystyle{jhep}
\bibliography{bkm}

\end{document}